\documentclass[]{aastex631}

\newcommand{\e}[1]{\ensuremath{\times 10^{#1}}}
\newcommand{\um}{$\mu$m }

\received{June 30, 2023}
\revised{September 7, 2023}
\accepted{September 8, 2023}

\submitjournal{PSJ}

\shorttitle{NH$_3$ and CO$_2$ on Miranda}
\shortauthors{DeColibus et al.}

\graphicspath{{./}{figures/}}

\begin{document}

\title{Are NH$_3$ and CO$_2$ ice present on Miranda?}

\correspondingauthor{Riley DeColibus}
\email{decolib@nmsu.edu}

\author[0000-0002-1647-2358]{Riley A. DeColibus}
\affiliation{Department of Astronomy, New Mexico State University \\
Box 30001, Dept.4500 \\
Las Cruces, NM 88003, USA}

\author[0000-0002-9984-4670]{Nancy J. Chanover}
\affiliation{Department of Astronomy, New Mexico State University \\
Box 30001, Dept.4500 \\
Las Cruces, NM 88003, USA}

\author[0000-0002-6886-6009]{Richard J. Cartwright}
\altaffiliation{Visiting Astronomer at the Infrared Telescope Facility, which is operated by the University of Hawaii under contract 80HQTR19D0030 with the National Aeronautics and Space Administration.}
\affiliation{The Carl Sagan Center at the SETI Institute\\
189 Bernardo Ave., Suite 200\\
Mountain View, CA 94043, USA}

\begin{abstract}
Published near-infrared spectra of the four largest classical Uranian satellites display the presence of discrete deposits of CO$_2$ ice, along with subtle absorption features around 2.2 \micron. The two innermost satellites, Miranda and Ariel, also possess surfaces heavily modified by past endogenic activity. Previous observations of the smallest satellite, Miranda, have not detected the presence of CO$_2$ ice, and a report of an absorption feature at 2.2 \um has not been confirmed. An absorption feature at 2.2 \um could result from exposed or emplaced NH$_3$- or NH$_4$-bearing species, which have a limited lifetime on Miranda's surface, and therefore may imply that Miranda's internal activity was relatively recent. In this work, we analyzed near-infrared spectra of Miranda to determine whether CO$_2$ ice and the 2.2-\um feature are present. We measured the band area and depth of the CO$_2$ ice triplet (1.966, 2.012, and 2.070 \micron), a weak 2.13-\um band attributed to CO$_2$ ice mixed with H$_2$O ice, and the 2.2-\um band. We confirmed a prior detection of a 2.2-\um band on Miranda, but we found no evidence for CO$_2$ ice, either as discrete deposits or mixed with H$_2$O ice. We compared a high signal-to-noise spectrum of Miranda to synthetic and laboratory spectra of various candidate compounds to shed light on what species may be responsible for the 2.2-\um band. We conclude that the 2.2-\um absorption is best matched by a combination of NH$_3$ ice with NH$_3$-hydrates or NH$_3$-H$_2$O mixtures. NH$_4$-bearing salts like NH$_4$Cl are also promising candidates that warrant further investigation. 
\end{abstract}


\section{Introduction and Background \label{sec:intro}}

Ground-based near-infrared spectroscopic observations of the five classical Uranian satellites have revealed that their surfaces are dominated by H$_2$O ice, mixed with a dark, low albedo component \citep{BrownCruikshank1983,BrownClark1984}. Ariel, Umbriel, Titania, and Oberon show spectral evidence of crystalline CO$_2$ ice, primarily concentrated on their trailing hemispheres \citep{Grundy2003,Grundy2006,Cart2015}. In contrast, Miranda does not show evidence of CO$_2$ ice \citep{Gourgeot2014,Cart2018}, possibly because its lower gravity prevents retention of CO$_2$ \citep[]{Grundy2006,Sori2017}. 

All five satellites show evidence of a weak absorption feature near 2.2-\um \citep{Bauer2002,Cart2018,Cart2020Ariel,Cart2023}. This 2.2-\um absorption feature appears qualitatively similar to the 2.21-\um band on Charon, which has been attributed to NH-bearing species such as NH$_3$-hydrates and NH$_4$Cl \citep[e.g.,][]{Cook2007,Cook2018,Cook2023}. Ammonia (NH$_3$) acts as an antifreeze and is predicted to have been incorporated into these icy bodies during their formation. Ammonia at the surface of icy bodies is thought to be dissociated by radiation on relatively short geological timescales \citep{Strazzulla1998,Moore2007}. Detection of NH-bearing species could then imply recent exposure by impacts, or perhaps emplacement by endogenic processes like cryovolcanism. Previous work noted a weak 2.2-\um band in spectra of Miranda \citep{Bauer2002}, but subsequent studies were unable to confirm its presence \citep{Gourgeot2014,Cart2018}. New ground-based near-IR spectra of Miranda were recently published with much higher signal-to-noise ratios (S/N) compared to prior studies \citep{DeColibus2022}. This new dataset was used to find subtle variations in H$_2$O ice band strengths between Miranda's anti-Uranus and sub-Uranus quadrants \citep{DeColibus2022}. We analyzed these higher quality spectra to perform a new, in-depth investigation to determine whether CO$_2$ ice and the 2.2-\um absorption feature are present on Miranda.

In the following subsections we describe the state of knowledge for the surface composition of Miranda and the other classical Uranian moons. We discuss our Miranda data set in Section \ref{sec:obs} and our analysis of integrated band areas and depths of the CO$_2$ ice triplet and the 2.13-\um and 2.2-\um absorption bands in Section \ref{sec:analysis}. We summarize our results in Section \ref{sec:results}, our spectral modeling in Section \ref{sec:modelresults}, and discuss the implications of our findings in Section \ref{sec:discussion}.

\subsection{Geology of Miranda\label{ssec:geology}}
\begin{figure}[ht!]
\centering
\plottwo{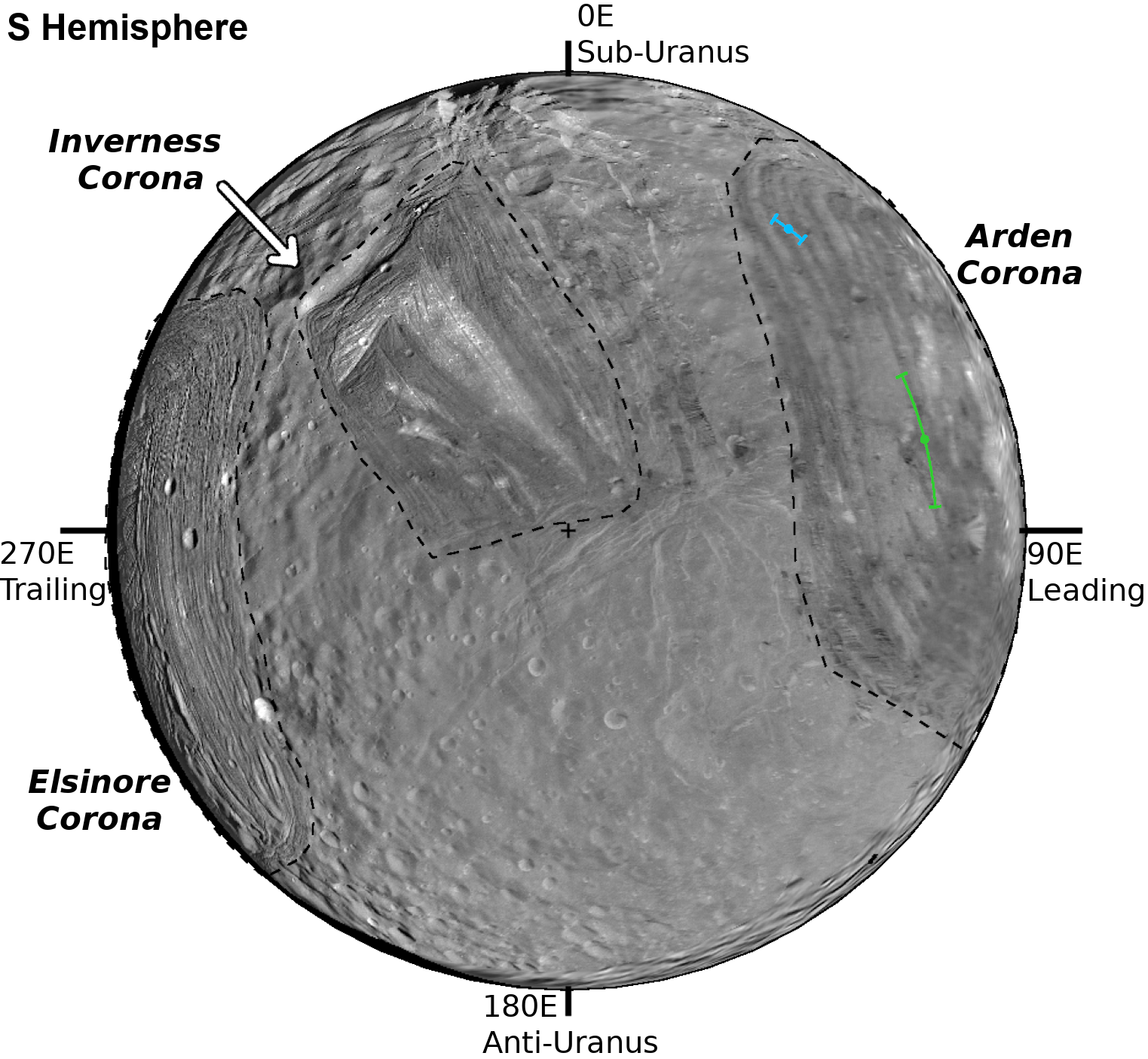}{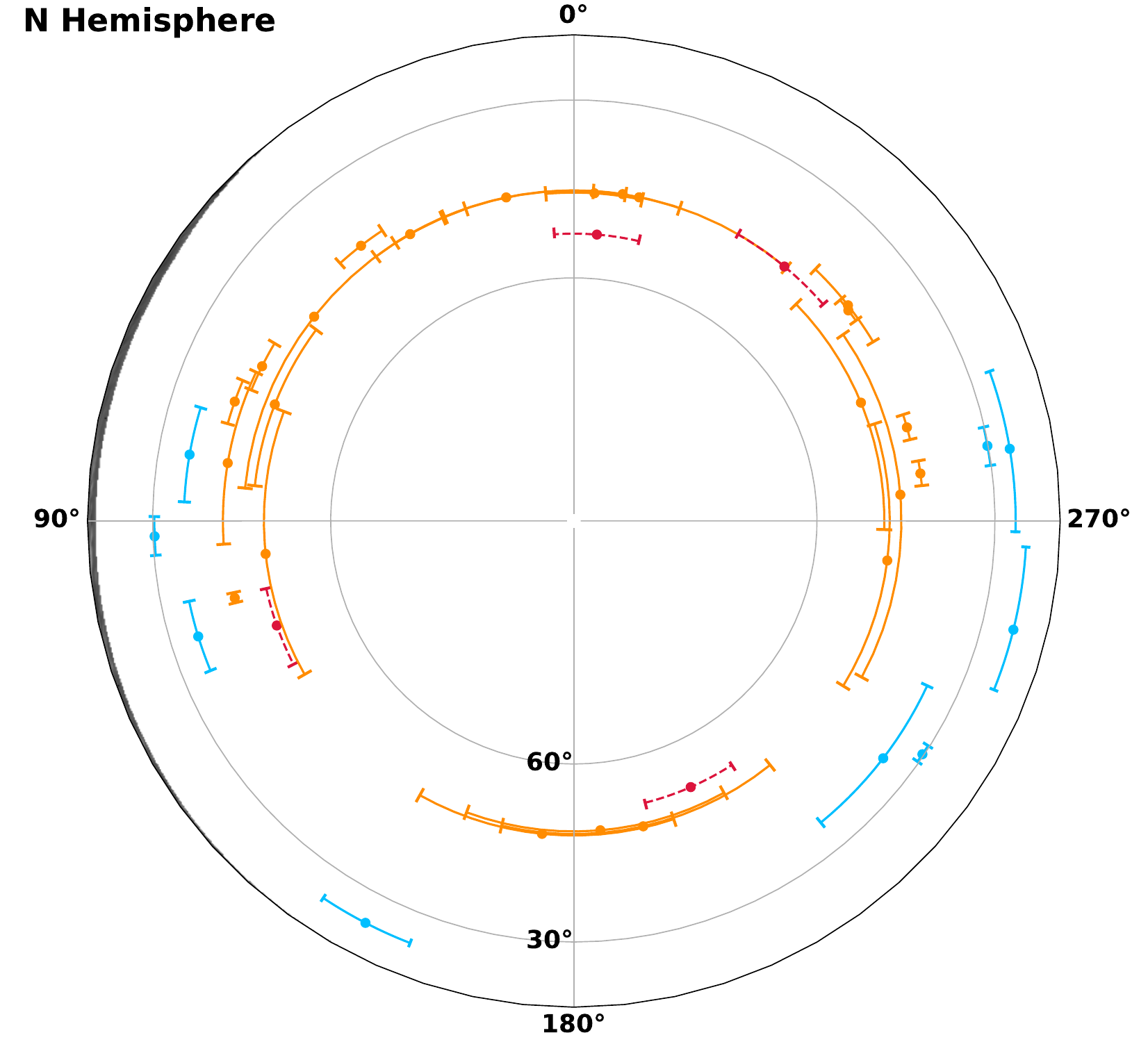}
\caption{\footnotesize (Left panel): An image mosaic of Miranda's southern hemisphere from imaging by Voyager 2 in 1986. We use an orthographic projection centered on the south pole, using the Miranda mosaic produced by \citet{Schenk2020}. The `coronae' are regions heavily modified by tectonic activity and possessing intriguing patches of high-albedo material. Only two of the disk-integrated spectra in our dataset were observed with sub-observer latitudes on the southern hemisphere: UT990607, green \citep{Bauer2002} and UT000907, blue \citep{Cart2018}. Errorbars in longitude represent the range of sub-observer longitudes covered during the duration of that observation. (Right panel): All other spectra in our dataset observed Miranda's northern hemisphere, which was in darkness at the time of the Voyager flyby and has no imaging data available. Orange spectra were observed with TripleSpec, blue spectra were observed with SpeX, and red spectra were observed with GNIRS. Longitudes appear to increase in opposite directions as seen from the north and south poles, so we chose to orient the 0\degr meridian at the top of the figure in both panels.}
\vspace{-5pt}
\label{fig:MirGeology}
\end{figure}

Miranda is the smallest of the five classical Uranian satellites. With a radius of $\sim$ 234 km, it is intermediate in size between the Saturnian moons Mimas and Enceladus. Prior to the Voyager 2 flyby of Uranus in 1986, Miranda was expected to possess a surface similar to Mimas (heavily cratered, with minimal to no evidence of geological activity). However, imaging of Miranda's southern hemisphere collected by Voyager 2 revealed an icy body where three large regions of its surface appeared to be geologically young (called ``coronae''), while other regions appeared to be ancient and heavily cratered (Figure \ref{fig:MirGeology})\citep{Smith1986,Greenberg1991,Schenk2020}. The coronae are bounded by large tectonic fault systems, and both Arden and Inverness Corona possess patches of high albedo material that might represent deposits of fresher H$_2$O ice. At the time, the incongruity between such a small icy body possessing complex tectonic and geological surface features led to the suggestion that Miranda had been disrupted into large chunks after a giant impact and subsequently reaccreted \citep{Smith1986,Janes1988}. Later work suggests that Miranda's bizarre coronae instead represent the surface expression of internal upwelling from low order convection \citep[e.g.][]{Papp1997,Hammond2014}, likely with tidal interactions from orbital resonances as a heat source. Crater counts suggest that the coronae are geologically young (0.1 -- 1 Gyr, \citet{Zahnle2003,Kirchoff2022}), and Miranda's larger neighbor Ariel also displays a relatively young surface, implying that Miranda and Ariel have both experienced complex geological histories. 

Furthermore, it has previously been noted that Miranda's large scale geology is reminiscent of the similarly sized active ocean world Enceladus, such as the presence of three regions of heavily tectonized terrain at the south pole and near the centers of the leading and trailing hemispheres, interspersed with heavily cratered ancient terrain \citep{Papp2013,Bedd2020}. Miranda is not currently in an orbital resonance that should produce significant internal heating, so the retention of an Enceladus-like subsurface ocean over geological timescales is unlikely \citep{Hussmann2006,CastilloRogez2023}. However, the surface geology indicates that Miranda likely experienced one or more significant heating events in the past \citep{Bedd2015,Schenk2020,Bedd2022}, and the apparent youthfulness of the coronae suggests that these events may have been geologically recent.

\subsection{\texorpdfstring{H$_2$O and CO$_2$}{H2O and CO2} ice}\label{ssec:CO2bkg}
The presence and spectral properties of H$_2$O ice on the Uranian satellites have been discussed at length in prior work \citep{BrownCruikshank1983,BrownClark1984,Grundy1999,Bauer2002,Grundy2003,Grundy2006,Cart2015,Cart2018,Cart2020IRAC,DeColibus2022}. We therefore give only a brief overview in this section. The surfaces of the Uranian satellites are characterized primarily by their lower albedos in comparison to similarly sized Saturnian satellites. These low albedos are attributed to a dark, spectrally neutral component (often assumed to be carbonaceous in nature) mantling or intermixed with the icy regolith. This dark material effectively erases the 1.04-\um and 1.25-\um H$_2$O ice bands and substantially decreases the depth of the 1.5-\um and 2.0-\um absorption band complexes. The four largest Uranian satellites show leading/trailing asymmetries in the strength of their H$_2$O ice bands, likely due to a combination of magnetospheric irradiation and impact gardening. Miranda does not show a leading/trailing asymmetry in the strength of its H$_2$O ice bands, although it possesses an anti-Uranus/sub-Uranus asymmetry instead \citep{Cart2018,DeColibus2022}.

Previous work reported the presence of CO$_2$ ice on the four largest Uranian satellites \citep{Grundy2003,Grundy2006,Cart2015,Cart2022}. This CO$_2$ ice is identified via a triplet of narrow absorption features at 1.966, 2.012, and 2.070 \micron. Weaker CO$_2$ ice absorption bands are also present at 1.543, 1.578, and 1.609 \micron, mostly on Ariel, which has the strongest CO$_2$ ice signatures. These CO$_2$ ice deposits are thought to be present primarily in pure deposits and not intermixed with the H$_2$O ice regolith. Furthermore, a forbidden $2\nu_3$ overtone mode of CO$_2$ exhibits an absorption feature near 2.134 \um \citep{Bernstein2005}, which only appears in molecular mixtures of CO$_2$ ice in H$_2$O ice and/or methanol (CH$_3$OH) ice, might be present on Ariel \citep{Cart2022} and Umbriel \citep{Cart2023}. 

The distribution of CO$_2$ ice on the Uranian moons shows both longitudinal and planetocentric trends. The CO$_2$ absorption features are stronger on moons closer to Uranus, and on each individual moon, the CO$_2$ ice bands are stronger on their trailing hemispheres. This information supports an interpretation in which CO$_2$ molecules are produced \textit{in situ} by magnetospheric bombardment of the trailing hemispheres of the Uranian satellites. Irradiation of the H$_2$O ice and carbonaceous compounds in the regolith dissociates molecules that then recombine into CO$_2$, potentially with CO as an intermediary product \citep{Cart2022}. Plasma densities in the magnetosphere are higher closer to Uranus, implying that a radiolytic production process of CO$_2$ should plausibly be more effective on moons closer to the planet. This assertion is supported by the relative strength of the CO$_2$ absorption features across the various satellites, as the CO$_2$ signature is strongest on Ariel and weakest on Oberon, which spends part of its orbit outside of the Uranian magnetosphere \citep{Ness1986}. Thermodynamical modeling work indicates that radiolytically generated CO$_2$ molecules should preferentially accumulate in cold traps at low latitudes on the trailing hemispheres of each satellite, assuming radiolytic production is greater on their trailing sides \citep{Grundy2006,Sori2017}. More recent telescope observations show a decrease in the strength of the CO$_2$ ice bands on Ariel at higher sub-observer latitudes, consistent with the hypothesis that CO$_2$ cold traps are concentrated at low latitudes \citep{Cart2022}.

Miranda is the innermost of the five classical satellites, and therefore, it should be the most irradiated by charged particles trapped in the Uranian magnetosphere. Miranda presumably possesses the same ``raw ingredients'' in its regolith (carbonaceous compounds and H$_2$O ice) as the other Uranian satellites \citep{BrownClark1984}, unless Miranda formed from a different mix of materials, possibly as a consequence of the Uranus-tilting event \citep{RufuCanup2022,SalmonCanup2022}. It therefore follows that radiolytic production of CO$_2$ molecules should also be occurring on Miranda, and it may also maintain deposits of CO$_2$ ice like the other Uranian moons. However, previous spectroscopic studies of Miranda have not detected CO$_2$ ice on its surface \citep{Bauer2002,Gourgeot2014,Cart2018}. These prior studies were somewhat limited by low signal-to-noise ratio (S/N) reflectance spectra in the 2.0-\um region, due to several observational limitations. Miranda is much fainter than the other Uranian moons at K$_{mag}\sim 15$, compared to K$_{mag}\sim$12-13 for the larger moons. The CO$_2$ ice absorption features of interest are narrow, requiring higher spectral resolving power and therefore lower S/N per spectral pixel. The CO$_2$ ice absorption bands are superimposed on the wide and deep 2.0-\um H$_2$O ice absorption band, and the CO$_2$ ice bands are also in a region of strong telluric absorption from CO$_2$ in the Earth's atmosphere. Although the spectral signature of atmospheric CO$_2$ is distinctly different than that of crystalline CO$_2$ ice \citep[e.g.,][]{Hansen1997}, telluric contamination introduces additional uncertainty into the spectral data points. These prior observations established CO$_2$ ice band depth upper limits of 5\% of the continuum on Miranda. Therefore, if CO$_2$ ice is present, it is much less abundant on Miranda than on Ariel \citep{Cart2018}. Additionally, CO$_2$ molecules in a molecular mixture with H$_2$O ice might be present and detectable prior to escaping Miranda's low gravity. Investigating whether Miranda spectra show the 2.13-\um $2\nu_3$ overtone band could help determine whether regolith-mixed CO$_2$ is present. The primary questions relating to CO$_2$ that this work aims to answer are therefore twofold: (1) do concentrated CO$_2$ ice deposits exist on Miranda's surface at low abundances, and (2) is there evidence for CO$_2$ in a molecular mixture with H$_2$O ice in the regolith?

\subsection{The \texorpdfstring{2.2-\um}{2.2-um} feature}\label{ssec:22umbkg}

Some icy bodies show evidence for weak absorption features in the region between 2.18 and 2.26 \micron, which we refer to generically as the 2.2-\um feature. The 2.2-\um feature on icy bodies has frequently been attributed to ammonia (NH$_3$)- or ammonium (NH$_4$)-bearing compounds (which we will collectively refer to as NH-bearing). NH-bearing species are of significant geological and astrobiological interest, as NH$_3$ is a potent antifreeze, capable of reducing the freezing point of liquid H$_2$O by nearly 100 K if present in sufficiently high abundances \citep{Kargel1992}. This makes NH$_3$ an important compound for enabling internal activity and extending the retention of subsurface oceans on icy bodies. NH$_4$-bearing salts and minerals are much less effective than NH$_3$ in an antifreeze role \citep{Neveu2017}, but indicate the past presence of NH$_3$, and NH$_4^+$ is readily produced by irradiation of H$_2$O + NH$_3$ ice mixtures \citep{Moore2003,Moore2007}.

Despite the expected ubiquity of NH$_3$-bearing compounds in the solar nebula at intermediate to large heliocentric distances \citep{Lewis1971}, the 2.2-\um feature is not observed in the spectra of many icy bodies, possibly because of removal by irradiation over short geological timescales. In the case of Miranda, NH$_3$ may be effectively removed from its surface in timescales as short as $<$10$^6$ years \citep{Moore2007}. Detection of the 2.2-\um feature could therefore imply that NH$_3$-bearing compounds have been exposed or emplaced in the geologically recent past, either via endogenic processes such as tectonism and cryovolcanism or via mass wasting and impact events. The possibility of recent endogenic activity on Miranda is particularly tantalizing given the apparent youth of some regions of its heavily modified surface, discussed in \S \ref{ssec:geology}, and past reports of a weak 2.2-\um absorption feature \citep{Bauer2002}.

\subsubsection{The \texorpdfstring{2.2-\um}{2.2-um} feature on other icy bodies}

One of the icy bodies that shows the most consistent evidence of a 2.2-\um feature is Charon \citep[e.g.,][]{BrownCalvin2000,Dumas2001}. Prior to the New Horizons flyby of the Pluto system, ground-based spectroscopic studies investigated the 2.2-\um feature on Charon in detail, finding evidence for longitudinal variation in the strength and wavelength position of the 2.2-\um band \citep{Cook2007,Merlin2010,DeMeo2015,Holler2017}. Data acquired during the New Horizons flyby further confirmed the band's presence on Charon and an apparent spatial association with impact craters \citep{Grundy2016,Cook2018,DalleOre2018}. \citet{DalleOre2019} and \citet{Cruikshank2019} found the 2.2-\um band to be present near geologically young surface features on Pluto that also exhibit spectral evidence of H$_2$O ice, strengthening a possible association with emplacement of NH-bearing compounds via cryovolcanism/internal activity. In contrast, \citet{Cook2018} reported on strong 2.2-\um bands on Pluto's minor moons Nix and Hydra, which are too small to support internal geological activity. 

In addition to the Uranian satellites and icy bodies in the Pluto system, detection of weak 2.2-\um features also have been reported on Enceladus \citep{Emery2005,Verb2006}, Tethys \citep{Verb2008}, Orcus \citep{Carry2011}, Haumea and its satellite Hi'iaka \citep{Barkume2006}, and Quaoar \citep{Jewitt2004}. NH$_4$-bearing species have been reported on Ceres \citep{King1992,DeSanctis2016,Raponi2019} and comets \citep{Poch2020}, and the 2.2-\um feature on Charon and the smaller moons in the Pluto system has been attributed to NH$_4$Cl \citep{Cook2018,Cook2023}. However, a 2.2-\um absorption band is not exclusive to NH-bearing species. Other minerals and compounds can also exhibit absorption in this region, including many phyllosilicates (hydrated silicates) with an Al-OH bond \citep{Clark1990Minerals}. Certain types of phyllosilicates are also prone to incorporating NH$_4$ into their composition through cation exchange processes \citep{Bishop2002,Berg2016}.

\subsubsection{The \texorpdfstring{2.2-\um}{2.2-um} feature on the Uranian satellites}

\citet{Bauer2002} was the first work to identify a 2.2-\um feature on any of the Uranian satellites, reporting a relatively strong band on Miranda (plotted in green in the left panel of Figure \ref{fig:CO2NH3vstack}). The authors reported an absorption band similar to that of the 2.21-\um feature on Charon, and attributed it to ammonia hydrate (NH$_3\cdot n$H$_2$O). At the time, Miranda was only the third icy body on which a 2.2-\um band had been reported. However, subsequent spectral studies of Miranda \citep{Gourgeot2014,Cart2018} did not find conclusive evidence of a 2.2-\um absorption feature, leaving its existence an open question. 

From visual inspection, \citet{Cart2018} noted weak absorption features in the 2.2-\um region on Ariel, Umbriel, Titania, and Oberon. The authors followed up with a detailed study of 2.2-\um features on Ariel in \citet{Cart2020Ariel}, finding several different absorption bands associated with NH$_3$- and NH$_4$-bearing compounds, which varied in strength and band center in different spectra. They found no large scale longitudinal patterns in these variations. They compared these bands with NH$_3$ hydrates (NH$_3\cdot n$H$_2$O, 2.215 \micron), flash frozen NH$_3$-H$_2$O solutions (2.210 \micron), NH$_3$ ice (2.238 \micron), and an NH$_4$-bearing species (possibly (NH$_4$)$_2$CO$_3$, 2.181 \micron). Some of these bands have been reported to show double absorptions and wavelength shifts based on temperature and the ratio of NH$_3$ to H$_2$O \citep{Moore2007}, further complicating the identification of individual compounds. 

Further research has also presented evidence of similar weak absorption features in the 2.2-\um region on Umbriel \citep{Cart2023}. However, the surface of Umbriel is ancient and heavily cratered (within the resolution limits of Voyager 2 imaging), lacking evidence of resurfacing driven by endogenic activity. The presence of 2.2-\um bands in spectra of Umbriel raises important questions about the presumed short lifetime of NH-bearing species on the surfaces of the Uranian moons \citep{Moore2007}, and also raises the possibility that other species such as phyllosilicate minerals or nitrogen-bearing organics contribute to the 2.2-\um band, at least on Umbriel.

\section{Observations} \label{sec:obs}

\begin{figure}[ht!]
\centering
\makebox[\textwidth][c]{\includegraphics[width=0.8\textwidth]{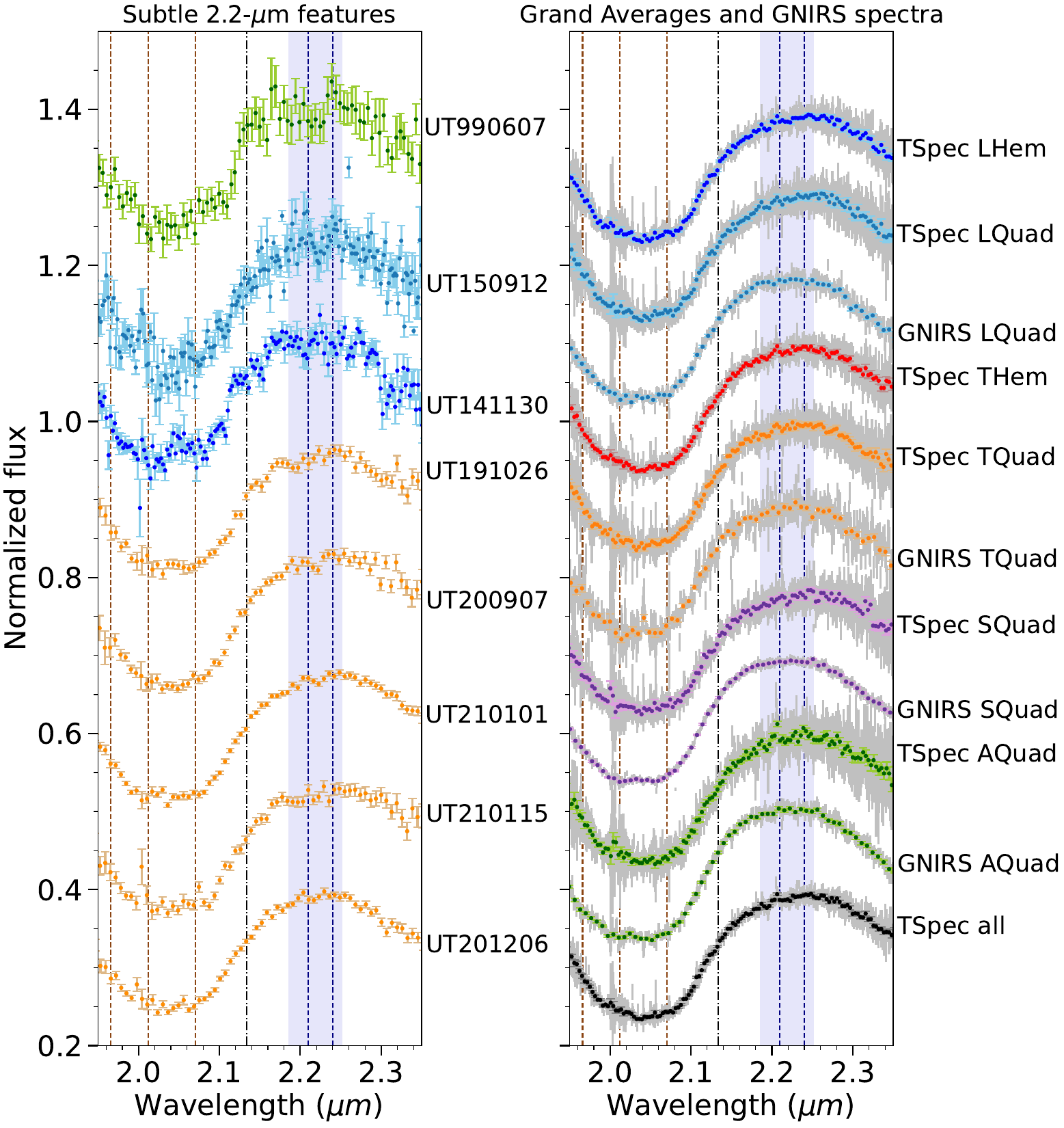}}
\vspace{-10pt}
\caption{\footnotesize (Left panel): Examples of Miranda spectra displaying weak absorption features around 2.2 \micron. The green spectrum is the (unbinned) spectrum reported in \citet{Bauer2002}. The two blue spectra were obtained with SpeX in the PRISM and SXD configuration (unbinned and binned by 10 pixels, respectively). The orange spectra are spectra from individual nights with TripleSpec (binned by 30 pixels). The text next to each spectrum indicates the date of observation in format UTYYMMDD (e.g. UT200907 = UT 2020-09-07). This UTYYMMDD date format is used throughout this work. Vertical lines are placed at the expected wavelengths for the absorption features studied in this work: brown lines at 1.966, 2.012, and 2.070 \um for the CO$_2$ ice triplet, a black line at 2.134 \um for a feature associated with molecular mixtures including CO$_2$ ice, a blue line at 2.21 \um for NH$_3$-hydrates, and a blue line at 2.24 \um for crystalline NH$_3$ ice. The shaded region indicates the wavelength range measured for the 2.2-\um band in this work. (Right panel): The GNIRS spectra and TripleSpec quadrant and hemisphere grand average spectra originally reported in \citet{DeColibus2022}. Colored points are binned by 15 pixels, and the gray error bars are presented at the native resolution of the data.}
\vspace{-5pt}
\label{fig:CO2NH3vstack}
\end{figure}

\begin{deluxetable*}{llcllccl}
\tabletypesize{\footnotesize}
\tablecaption{Miranda Observations\label{tab:obstable}}
\tablehead{
\colhead{Observing PI} & \colhead{Year(s)} & \colhead{Latitude(s)} &
\colhead{Telescope}& \colhead{Instrument} & \colhead{Resolving power} & \colhead{N$_{spec}$} & \colhead{Reference} \\
\colhead{} & \colhead{} & \colhead{\degr N, Sub-Earth} & \colhead{} & \colhead{} &
\colhead{($\lambda/\Delta\lambda$)} & \colhead{} & \colhead{}
}  
\startdata
Bauer & 1999 & -36.5 & UKIRT & CGS4 & $\sim$200 & 1 & \citet{Bauer2002} \\
Rivkin & 2000 & -35.4 & IRTF & SpeX SXD & $\sim$750 & 1 & \citet{Cart2018} \\
Gourgeot & 2012 & 21.4 & IRTF & SpeX SXD & $\sim$750 & 2 & \citet{Gourgeot2014} \\
Cartwright & 2014--2017 & 24.7 -- 36.7 & IRTF & SpeX SXD & $\sim$750 & 2 & \citet{Cart2018} \\
Cartwright & 2014--2017 & 24.7 -- 36.7 & IRTF & SpeX PRISM & $\sim$95 & 5 & \citet{Cart2018} \\
DeColibus & 2019--2021 & 42.3 -- 50.4 & ARC 3.5m & TripleSpec & $\sim$3500 & 18 & \citet{DeColibus2022} \\
DeColibus & 2020 & 47.2 -- 49.6 & Gemini N & GNIRS XD & $\sim$1130 & 2 & \citet{DeColibus2022} \\
DeColibus & 2021 & 53.3 -- 53.8 & Gemini N & GNIRS XD & $\sim$750 & 2 & \citet{DeColibus2022} \\
\enddata
\tablecomments{\footnotesize A table of basic information about the collection of Miranda spectra analyzed in this work. For more details, we direct the reader to the listed references.}
\end{deluxetable*}

We analyzed 33 disk-integrated near-IR spectra of Miranda, covering the wavelengths between $\sim$0.95 -- 2.45 \micron, summarized in Table \ref{tab:obstable}. Eighteen of these spectra were obtained with the TripleSpec spectrograph on the ARC 3.5-meter telescope at Apache Point Observatory, four spectra were acquired with GNIRS on the 8.1-meter Gemini North telescope on Maunakea, and ten spectra were observed with SpeX on NASA's 3-meter Infrared Telescope Facility (IRTF) on Maunakea. For information on the acquisition and data reduction of the TripleSpec and GNIRS spectra, we refer the reader to \citet{DeColibus2022}, while for the SpeX spectra, we refer the reader to \citet{Gourgeot2014} and \citet{Cart2018}. Table \ref{tab:obstable} lists the average spectral resolving power ($R = \lambda/\Delta\lambda$) of data acquired with each instrument configuration. Two spectral data points can be placed across the narrow CO$_2$ absorption bands at R$\sim$750, so they should be detectable in spectra acquired at this or higher spectral resolution. All of the spectra except those acquired with the SpeX PRISM configuration have sufficient spectral resolution to at least detect the CO$_2$ bands, and the 2.2-\um feature is broad enough to be detectable in the PRISM spectra. The dataset also includes seven TripleSpec `grand average' (GAvg) spectra, which were constructed as averages of all TripleSpec exposures in which the sub-observer longitude fell within defined ranges. These are the leading, trailing, sub-Uranus, and anti-Uranus quadrants; the leading and trailing hemispheres; and all exposures regardless of longitude. These grand average spectra and the GNIRS spectra are plotted in Figure \ref{fig:CO2NH3vstack}.

We also modified one spectrum in the above dataset. The GNIRS anti-Uranus quadrant spectrum from UT211121 displayed a high S/N, but the error bars on the spectrum were uniformly overestimated compared to what might reasonably be expected given the apparent S/N of the data and when compared to the other GNIRS spectra obtained in similar conditions (see Figure 5 of \citet{DeColibus2022}). In order to calculate more accurate uncertainties for each data point while retaining wavelength-dependent error information, such as increased uncertainties in regions of heavy telluric absorption, we adopted a procedure similar to that of \citet{Holler2022}. We calculated the standard deviation of the flux values and the mean of the original uncertainty values in a region of the spectrum that could reasonably be described by a flat continuum (1.70 -- 1.80 \micron). Division of these two quantities results in the factor by which the errors in the original spectrum were overestimated, approximately $\sim$3.32. Division by this factor resulted in a more accurate estimate of the uncertainties for the data points in that spectrum, and the corrected spectrum was used for all analyses presented here. 

Finally, we also digitized the spectrum of Miranda's leading hemisphere published in \citet{Bauer2002}, in which the presence of a 2.2-\um feature was first reported. The K-band portion of this spectrum was acquired on 1999 June 7 with the CGS4 spectrograph on the 3.8-meter UKIRT telescope on Maunakea. We utilized timing and observing geometry information provided in Table 1 of \citet{Bauer2002}, combined with the JPL HORIZONS ephemeris service, to derive a central sub-observer longitude of 75.7\degr E and a latitude of 36.5\degr S. For further information on the Bauer et al. spectrum, we direct the reader to the original work. This spectrum and one SpeX SXD spectrum acquired in 2000 at latitude 35.4\degr S are the only two southern hemisphere spectra in our data set. All other spectra were acquired at northern sub-observer latitudes (Figure \ref{fig:MirGeology}). 

\section{Spectral Analysis\label{sec:analysis}}
We measured the integrated band areas of three CO$_2$ ice bands in the Miranda spectra. The CO$_2$ ice bands at 1.966, 2.012, and 2.070 \um are referred to in this work as the ``CO$_2$ ice triplet'' as a whole and as bands 1, 2, and 3 individually. We also measured the integrated band areas and fractional band depths of a subtle feature at 2.13-\um and the 2.2-\um absorption band.

\subsection{Band measurement methods}\label{ssec:measmethods}
\begin{figure}[ht!]
\centering
\makebox[\textwidth][c]{\includegraphics[width=0.7\textwidth]{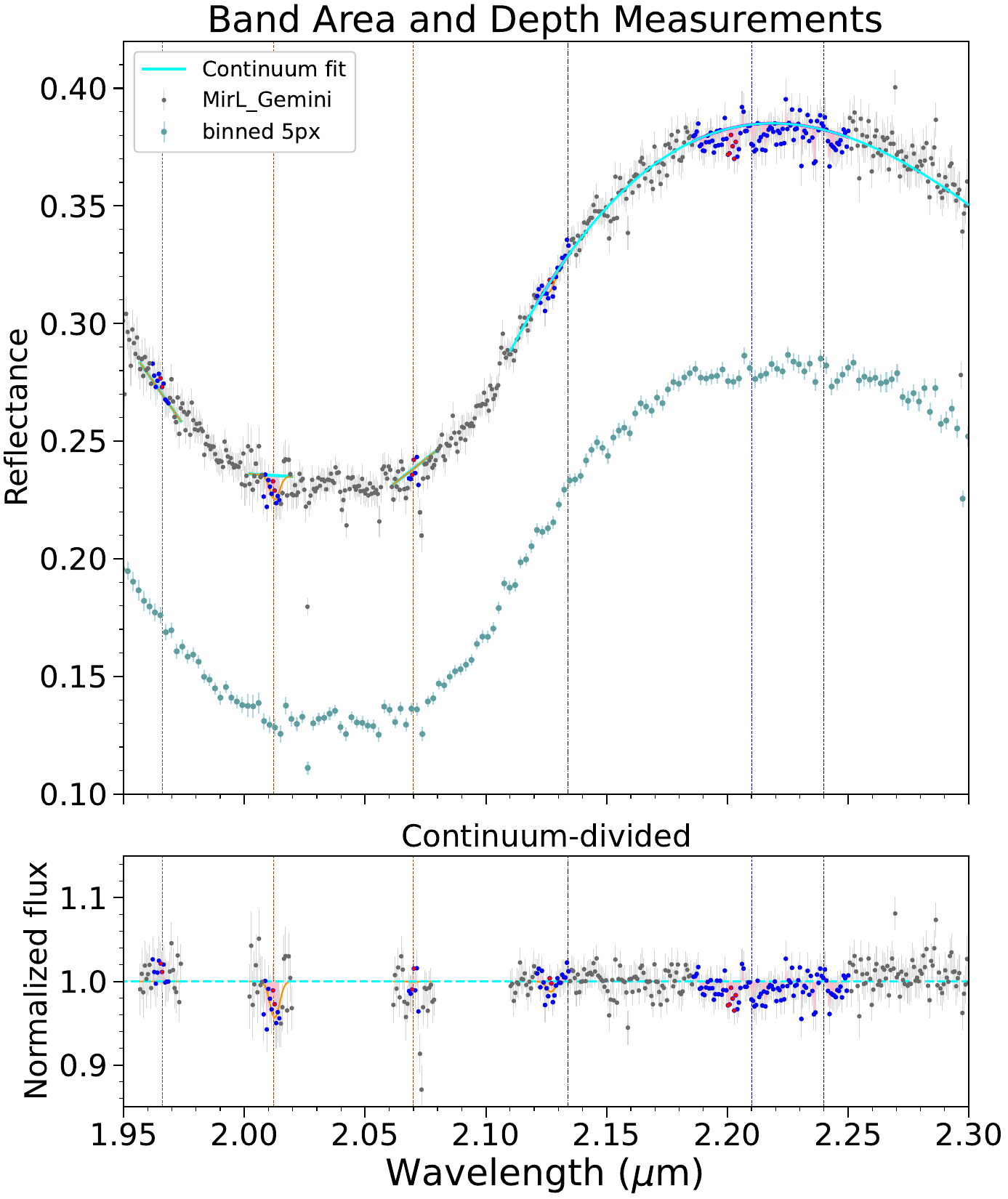}}
\caption{\footnotesize A demonstration of the band area and depth measurement process for the CO$_2$ ice triplet and the 2.13-\um and 2.2-\um bands, using the GNIRS leading hemisphere spectrum of Miranda from UT201008. The data points over which the band areas are integrated are marked in blue, the points where the band depth is measured in red, continuum fits are cyan lines, and Gaussian band fits are orange lines. A copy of the spectrum plotted with an offset has been binned by five pixels (turquoise points). Vertical lines are the same as in Figure \ref{fig:CO2NH3vstack}.}
\label{fig:banddefinitions}
\end{figure}

\begin{deluxetable}{ccccc}
\tablecaption{Absorption band wavelengths\label{tab:bandwaves}}
\tablehead{\colhead{Band name} & \colhead{Left continuum} & \colhead{Band width} & \colhead{Right continuum} & \colhead{Center}\\ 
\colhead{} & \colhead{($\mu m$)} & \colhead{($\mu m$)} & \colhead{($\mu m$)} & \colhead{($\mu m$)} }
\startdata
CO$_2$ 2$\nu_1$+$\nu_3$ (Band 1) & 1.957 -- 1.962 & 1.962 -- 1.969 & 1.969 -- 1.974 & 1.966 \\
CO$_2$ $\nu_1$+2$\nu_2$+$\nu_3$ (Band 2) & 2.002 -- 2.008 & 2.008 -- 2.015 & 2.015 -- 2.020 & 2.012 \\
CO$_2$ 4$\nu_2$+$\nu_3$ (Band 3) & 2.062 -- 2.068 & 2.068 -- 2.073 & 2.073 -- 2.079 & 2.070 \\
2.13-\um band & \nodata & 2.121 -- 2.135 & \nodata & 2.134 \\
2.2-\um band & \nodata & 2.186 -- 2.251 & \nodata & 2.21 \\
\enddata
\tablecomments{\footnotesize We tabulate the wavelengths we used to define the band continua, widths, and centers. For the 2.13-\um and 2.2-\um band, the continuum was fit with the 3rd degree polynomial fit described in the text. `Center' refers to the expected central wavelength of the absorption band.}

\end{deluxetable}
Our band area and depth measurements were conducted with the same analysis routine described in \citet{DeColibus2022}, modified for use on the CO$_2$ ice bands, 2.13-\um band, and 2.2-\um band. This routine is a Python implementation of the technique described in \citet{Cart2015,Cart2018}, which used a modified version of the SARA band analysis routine originally developed for asteroid spectra \citep{Lindsay2015}. The band analysis code generates a sample spectrum, drawing from a Gaussian distribution where the mean and standard deviation are the flux and errors of the input spectrum, respectively. The absorption bands of interest (Table \ref{tab:bandwaves}) are measured by drawing a line between the continuum on either side of the absorption band, normalizing by this line, and integrating the area inside the absorption band with the trapezoidal rule. The CO$_2$ ice absorption bands are further fit by a Gaussian function to determine the band centers, from which the depths of the bands are measured using the points within 0.0005 \um of the center of the band. This choice of band width for the depth measurement was determined from inspection of the CO$_2$ ice bands in spectra of Ariel. This process is demonstrated visually using a GNIRS spectrum of Miranda's leading hemisphere in Figure \ref{fig:banddefinitions}. To calculate the uncertainties on each measurement, this entire sample spectrum generation and measurement process is repeated 20,000 times, which is a number of samples consistent with other Monte Carlo type approaches to measurements of band parameters \citep[e.g.][]{Lindsay2015,Cart2015}. The mean and standard deviation of the measurements of this ensemble are reported as the final band measurements for each individual spectrum in Table \ref{tab:CO2bands}. For each spectrum, we also summed the band area measurements for the three CO$_2$ bands into a total CO$_2$ band area. Due to inadequate spectral resolution, we did not measure the CO$_2$ ice triplet for the \citet{Bauer2002} spectrum nor the SpeX spectra collected in PRISM mode.

The process for the 2.13-\um and 2.2-\um bands was slightly different. Instead of measuring the continuum by drawing a line on either side of the absorption band, the continuum is defined with a third-order polynomial curve, fit to the data points between 2.11 -- 2.19 \um and 2.23 -- 2.34 \micron. The region between 2.19 -- 2.23 \um is not used in the fit to avoid the influence of any possible 2.2-\um absorption features. The spectrum is divided by this continuum model, and the integrated band areas and fractional band depths are measured in wavelength ranges corresponding to the absorption bands of interest. We used a Gaussian fit procedure similar to the one used for the CO$_2$ ice triplet to measure the central wavelength and depth of the 2.13-\um band. However, for the 2.2-\um band, the potential presence of multiple absorptions across the 2.18 -- 2.25 \um range would be poorly fit by a Gaussian model. We instead binned the spectrum by five pixels and chose the lowest data point in the continuum-divided band as the central wavelength. The fractional band depth for the 2.2-\um band are measured using the mean of the unbinned, native resolution data points within $\pm$0.002 \um of this central wavelength. This process was also iterated 20,000 times for each spectrum, and the mean and standard deviation of the measurements for each individual spectrum are reported in Table \ref{tab:nh3bands}. 

\subsection{Means, ratios, and sinusoidal models}
We calculated the mean band parameters, using the final measurements of each of the spectra that fall within given sub-observer longitude quadrants and hemispheres. We also calculated a set of mean measurements that included all spectra from individual nights (i.e. not including the TripleSpec grand averages). These mean measurements are tabulated in Table \ref{tab:meanbands} and plotted in Figure \ref{fig:meanbandareas}. Figure \ref{fig:meanbandareas} also includes the band measurements from the quadrant and hemisphere-averaged TripleSpec grand average spectra and the GNIRS spectra for comparison to the mean measurements.

We took the ratios of the mean band measurements between opposing quadrants and hemispheres (Table \ref{tab:ratiobands}) to test whether a band was statistically stronger on one quadrant/hemisphere versus the other, as might be expected from exogenic effects. We did not find statistically significant ($\geq 2\sigma$) departures from a ratio of unity for any of the measurements or quadrant/hemisphere pairs.

We fit our band measurements to a sinusoidal model to search for longitudinal variation, with the period fixed at 2$\pi$ to represent one rotation of the body. This approach has previously been used for Miranda and the other Uranian satellites \citep{Grundy2006,Cart2015,Cart2018,DeColibus2022} and we direct the reader to those works for further information. We also applied an F-test to the sinusoidal models, which allows us to discern whether the sinusoidal model fit is a statistically significant better fit compared to the null hypothesis (no variation with sub-observer longitude). The sinusoidal models and F-tests are tabulated in Table \ref{tab:Ftable}.

\subsection{Hapke-Mie modeling \label{ssec:hapkeintro}}
We constructed spectral models in order to investigate possible absorption features in the 2.2-\um region. We chose a hybrid Hapke-Mie model using intimate (particulate) mixing, as has previously been done for the Uranian satellites \citep{Cart2015,Cart2018,Cart2020IRAC}. The Hapke-Mie model utilizes Mie theory to calculate the single scattering albedo of individual particles, which are then incorporated into the standard Hapke model equations. We assume an isotropic single scattering phase function which does not depend on wavelength. The Hapke-Mie approach accounts for the effects of grain sizes approaching wavelength, such as diffraction and Rayleigh scattering \citep{Clark2012}. However, at some particle sizes the exact Mie solution can introduce periodic resonance artifacts in the resulting spectrum. We mitigate this effect by calculating the average albedo of a spread of sizes around the specified grain size, typically with a 10\% spread in particle diameter. 

As has been noted by previous studies, Miranda's spectrum is consistent with a surface composition incorporating H$_2$O ice and a dark, spectrally neutral component, such as amorphous carbon \citep{BrownClark1984,Bauer2002,Gourgeot2014}. We scaled our observed spectra to a geometric albedo of 0.434 at 1.72 \um (from Figure 7 of \citet{Kark2001}). For our models, we used the optical constants of crystalline H$_2$O ice at 80 K from \citet{Mastrapa2008} and optical constants of amorphous carbon (sample BE1) from \citet{RouleauMartin1991}. 

We found that a small percentage of intimately mixed sub-micron H$_2$O ice grains assist in reproducing Miranda's spectrum. The central wavelength of the 2.0-\um H$_2$O ice band is shifted to slightly longer wavelengths, the continuum in the 2.2-\um region is low compared to the 1.8-\um continuum, and the blue slope from 2.3 -- 2.5 \um is steeper than in typical H$_2$O ice models of larger diameter grains. These characteristics are consistent with the presence of sub-micron H$_2$O ice grains \citep{Clark2012}, and previous studies have suggested that the regoliths of the Uranian satellites show a `fluffy' structure and evidence of the presence of tiny grains with sub-micron diameters, likely composed of H$_2$O ice \citep{Afanasiev2014,Cart2020IRAC}.

We adjusted individual parameters (grain size and mixing ratio) of the model components until we found a fit that generally reproduced the continuum shape of our spectra between 2.0 -- 2.4 \micron. We specifically attempted to be agnostic regarding any possible absorption features in the 2.2-\um region, as the possible presence of several overlapping weak bands made identifying a `clean' continuum difficult. because of degeneracies between model parameters, our spectral models provide useful but non-unique solutions. We also experimented with the addition of small amounts of other compounds (such as NH$_3$ ice) to study their effects on the resulting spectra. This analysis is described further in \S \ref{sec:modelresults}.

\section{Results}\label{sec:results}

\begin{figure}[ht!]
\centering
\makebox[\textwidth][c]{\includegraphics[width=0.9\textwidth]{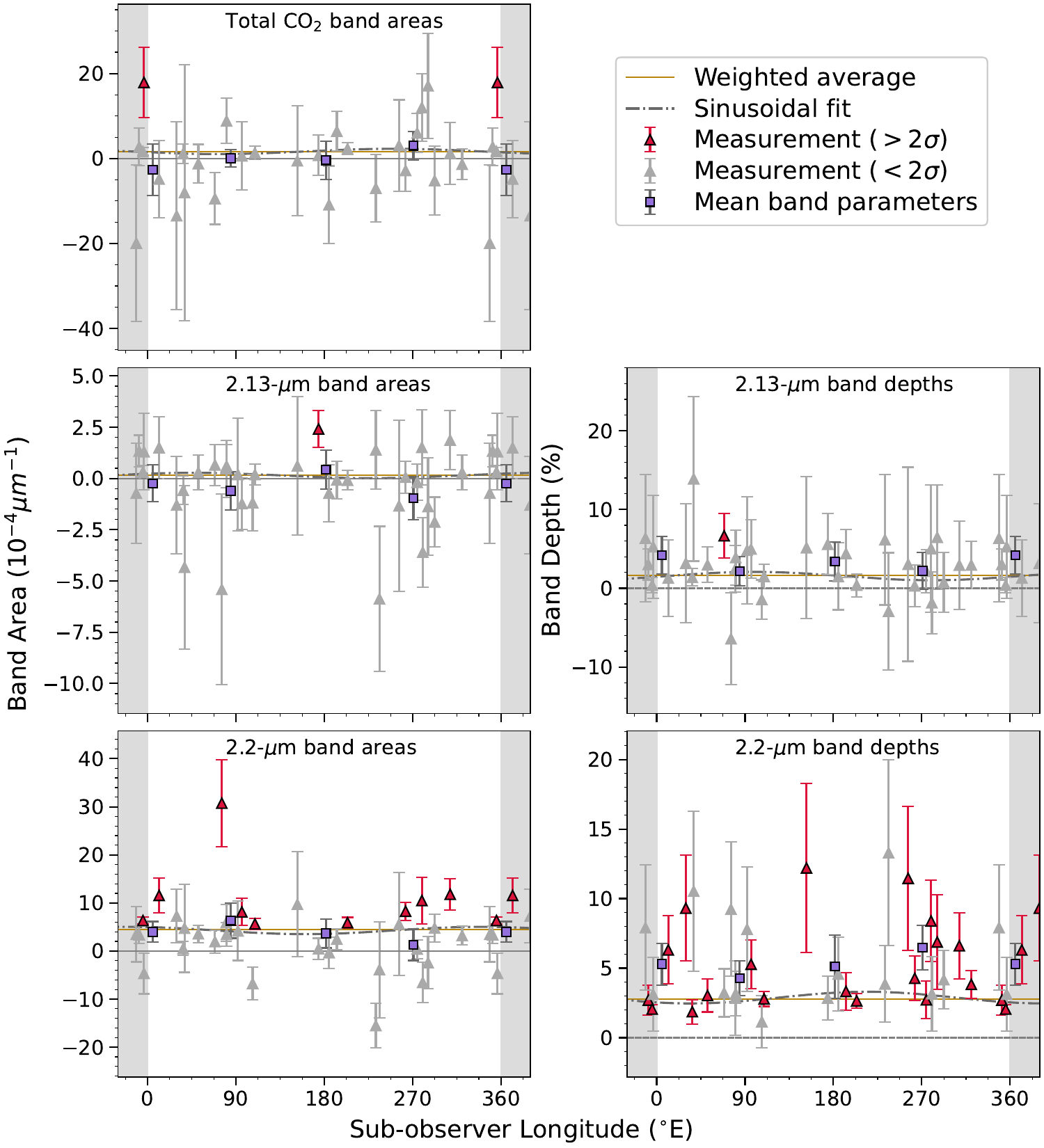}}
\vspace{-15pt}
\caption{\footnotesize Band area and depth measurements for each Miranda spectrum in this work, with 1$\sigma$ error bars. Band area and depth measurements that are not statistically significant ($<$2$\sigma$ from zero) are plotted as gray triangles, while those that are $\geq$2$\sigma$ are plotted as red triangles. Quadrant-averaged mean band parameters (Table \ref{tab:meanbands}) are plotted as purple squares. The gray shaded regions on either side of each plot represent overlapping longitudes. Data points are duplicated in these regions to better visualize possible sinusoidal variations with longitude. The weighted average and sinusoidal fit to all spectra for each measurement are plotted as golden horizontal lines and gray dot-dashed lines, respectively. Top left: Total band areas for the CO$_2$ ice triplet measurements (sum of the 1.966, 2.012, and 2.070 \um band areas). Center row: Band area measurements (left panel) and band depths (right panel) for the 2.13-\um band. Bottom row: Same as the center row, but for the 2.2-\um band. }
\vspace{-5pt}
\label{fig:bandareas}
\end{figure}

\begin{figure}[ht!]
\centering
\makebox[\textwidth][c]{\includegraphics[width=0.9\textwidth]{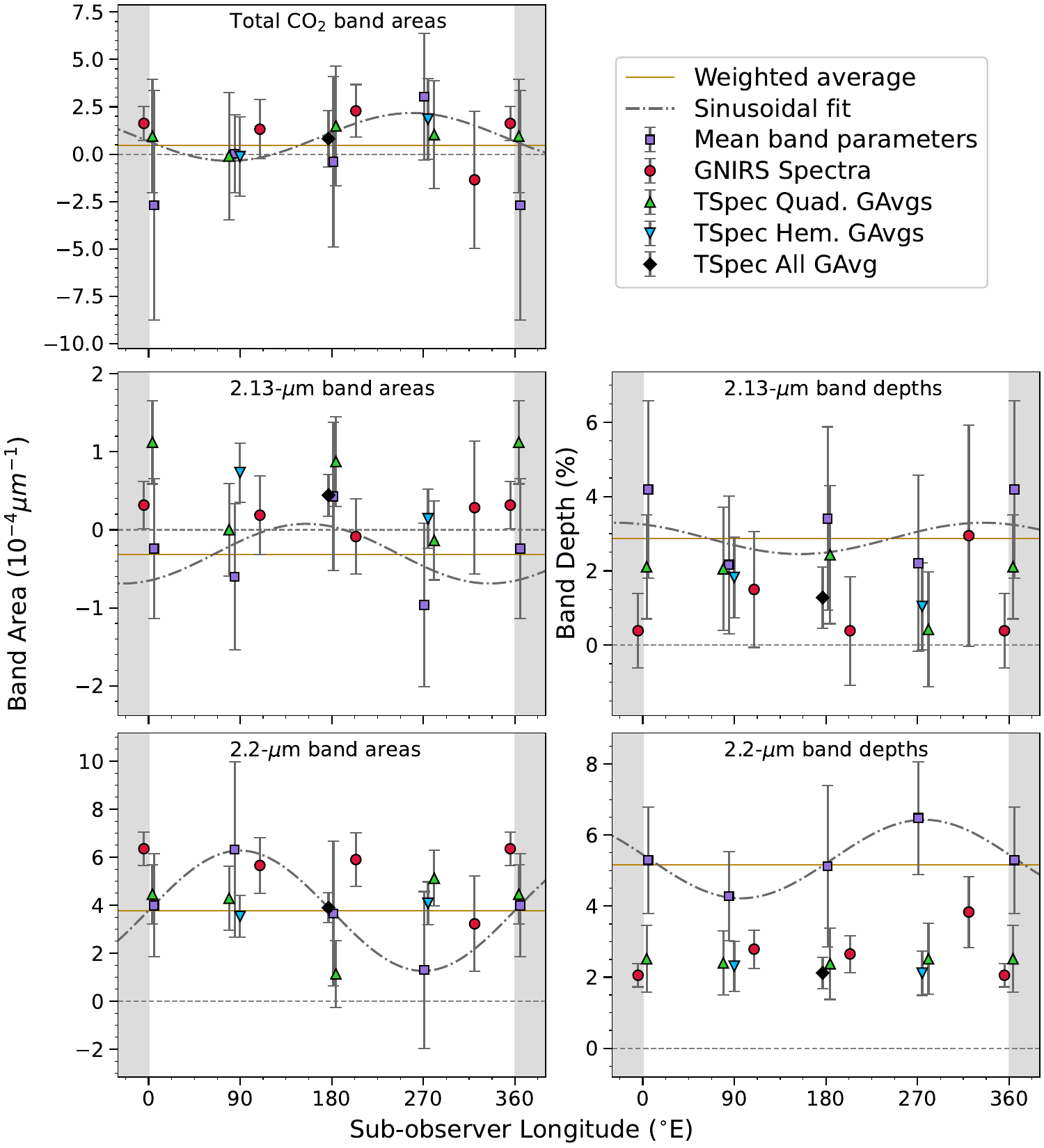}}
\vspace{-15pt}
\caption{\footnotesize Band area and depth measurements for the high S/N Miranda spectra (GNIRS spectra and the TripleSpec quadrant/hemisphere grand average spectra), along with the same mean measurements (purple squares) plotted in the previous figure. All error bars represent 1$\sigma$ uncertainties. The weighted average and sinusoidal fits use the mean band measurements.}
\vspace{-5pt}
\label{fig:meanbandareas}
\end{figure}

\begin{deluxetable}{ccccccccccc}
\tabletypesize{\footnotesize}
\tablecaption{CO$_2$ triplet band measurements\label{tab:CO2bands}}
\tablehead{\colhead{UT Date} & \colhead{Long.} & \colhead{Lat.} & \multicolumn{3}{c}{Integrated band area} & \colhead{Total CO$_2$} & \colhead{$>2\sigma$} & \multicolumn{3}{c}{Band depths} \\ 
\cline{4-6} \cline{9-11}
\colhead{} & \colhead{} & \colhead{} & \colhead{1.966-$\mu$m} & \colhead{2.012-$\mu$m} & \colhead{2.070-$\mu$m} & \colhead{band area} & \colhead{total} & \colhead{1.966-$\mu$m} & \colhead{2.012-$\mu$m} & \colhead{2.070-$\mu$m} \\
\colhead{} & \colhead{(\degr E)} & \colhead{(\degr N)} & \colhead{(10$^{-4}\mu$m)} & \colhead{(10$^{-4}\mu$m)} & \colhead{(10$^{-4}\mu$m)} & \colhead{(10$^{-4}\mu$m)} & \colhead{area?} & \colhead{(\%)} & \colhead{(\%)} & \colhead{(\%)}}
\startdata
UT210115 & 11.8 & 47.2 & $-2.32\pm5.07$ & $-2.60\pm7.33$ & $0.08\pm1.97$ & $-4.84\pm9.13$ & No & $-0.00\pm0.12$ & $0.01\pm0.16$ & $0.04\pm0.08$ \\
UT201229b & 29.7 & 47.2 & $-11.33\pm13.59$ & $-1.76\pm17.21$ & $-0.39\pm3.24$ & $-13.48\pm22.17$ & No & $-0.07\pm0.27$ & $0.23\pm0.32$ & $0.09\pm0.12$ \\
UT000907 & 36.3 & -35.4 & $0.51\pm1.07$ & $-0.00\pm1.91$ & $0.64\pm0.59$ & $1.15\pm2.27$ & No & $0.02\pm0.03$ & $0.07\pm0.04$ & $0.03\pm0.04$ \\
UT191022a & 37.7 & 44.3 & $-7.37\pm14.71$ & $-6.94\pm25.97$ & $6.28\pm4.22$ & $-8.03\pm30.15$ & No & $0.08\pm0.39$ & $0.00\pm0.50$ & $0.22\pm0.16$ \\
UT210101 & 51.8 & 47.2 & $3.31\pm2.02$ & $-3.14\pm3.93$ & $-1.41\pm1.10$ & $-1.23\pm4.55$ & No & $0.08\pm0.05$ & $0.01\pm0.10$ & $-0.01\pm0.04$ \\
UT201104 & 68.7 & 48.7 & $-0.75\pm3.21$ & $-7.18\pm4.97$ & $-1.51\pm1.38$ & $-9.44\pm6.07$ & No & $0.08\pm0.07$ & $-0.06\pm0.10$ & $-0.05\pm0.05$ \\
UT990607 & 75.7 & -36.5 & \nodata & \nodata & \nodata & \nodata & No & \nodata & \nodata & \nodata \\
UT170930 & 80.2 & 36.7 & \nodata & \nodata & \nodata & \nodata & No & \nodata & \nodata & \nodata \\
UT191104 & 80.5 & 43.8 & $1.84\pm3.57$ & $\mathbf{7.66\pm3.72}$ & $-0.66\pm1.38$ & $8.84\pm5.34$ & No & $0.08\pm0.11$ & $0.13\pm0.12$ & $0.00\pm0.05$ \\
UT150912 & 92.1 & 30.4 & \nodata & \nodata & \nodata & \nodata & No & \nodata & \nodata & \nodata \\
UT200907 & 96.1 & 50.4 & $2.12\pm5.72$ & $2.38\pm5.36$ & $-3.87\pm1.85$ & $0.63\pm8.06$ & No & $0.07\pm0.12$ & $0.13\pm0.11$ & $-0.04\pm0.08$ \\
UT171010 & 107.1 & 36.1 & \nodata & \nodata & \nodata & \nodata & No & \nodata & \nodata & \nodata \\
UT201008g & 109.4 & 49.6 & $-0.70\pm0.98$ & $1.84\pm1.11$ & $0.17\pm0.48$ & $1.31\pm1.56$ & No & $-0.01\pm0.02$ & $0.03\pm0.03$ & $0.01\pm0.03$ \\
UT120926 & 152.6 & 21.4 & $-6.72\pm8.24$ & $6.31\pm9.35$ & $-0.14\pm3.53$ & $-0.55\pm12.96$ & No & $-0.01\pm0.19$ & $0.14\pm0.21$ & $0.06\pm0.13$ \\
UT201007 & 174.2 & 49.7 & $1.88\pm2.56$ & $-1.10\pm3.87$ & $-0.03\pm1.15$ & $0.75\pm4.78$ & No & $0.03\pm0.07$ & $0.05\pm0.10$ & $0.00\pm0.05$ \\
UT200913 & 184.9 & 50.3 & $-3.33\pm5.43$ & $-5.83\pm7.08$ & $-1.73\pm1.94$ & $-10.89\pm9.13$ & No & $-0.02\pm0.12$ & $-0.04\pm0.17$ & $0.02\pm0.07$ \\
UT200930 & 192.8 & 49.9 & $2.79\pm2.50$ & $3.53\pm3.78$ & $0.09\pm1.21$ & $6.41\pm4.69$ & No & $0.07\pm0.06$ & $0.11\pm0.08$ & $0.03\pm0.05$ \\
UT211121g & 203.7 & 53.3 & $0.45\pm0.82$ & $1.37\pm1.00$ & $0.46\pm0.52$ & $2.28\pm1.39$ & No & $0.01\pm0.02$ & $0.04\pm0.02$ & $0.02\pm0.02$ \\
UT170925 & 232.5 & 36.7 & $2.16\pm3.18$ & $-6.70\pm6.83$ & $-2.47\pm2.36$ & $-7.01\pm7.89$ & No & $0.13\pm0.09$ & $-0.01\pm0.17$ & $-0.03\pm0.11$ \\
UT150911 & 236.2 & 30.4 & \nodata & \nodata & \nodata & \nodata & No & \nodata & \nodata & \nodata \\
UT120925 & 256.1 & 21.4 & $1.04\pm6.31$ & $1.87\pm7.57$ & $0.11\pm4.34$ & $3.02\pm10.77$ & No & $0.09\pm0.15$ & $0.07\pm0.17$ & $0.06\pm0.15$ \\
UT201013 & 262.8 & 49.5 & $2.59\pm2.51$ & $-5.23\pm4.03$ & $-0.19\pm1.03$ & $-2.83\pm4.86$ & No & $0.05\pm0.05$ & $-0.06\pm0.09$ & $0.03\pm0.04$ \\
UT201206 & 274.6 & 47.7 & $1.79\pm2.80$ & $5.01\pm3.46$ & $-0.80\pm1.14$ & $6.00\pm4.60$ & No & $0.03\pm0.06$ & $0.10\pm0.08$ & $-0.02\pm0.05$ \\
UT141130 & 279.4 & 24.7 & $6.22\pm3.80$ & $3.88\pm6.61$ & $1.89\pm2.40$ & $11.99\pm8.00$ & No & $0.11\pm0.15$ & $0.14\pm0.16$ & $0.06\pm0.15$ \\
UT150917 & 280.3 & 30.2 & \nodata & \nodata & \nodata & \nodata & No & \nodata & \nodata & \nodata \\
UT191013a & 285.7 & 44.6 & $3.95\pm8.43$ & $8.51\pm8.52$ & $4.64\pm3.15$ & $17.10\pm12.39$ & No & $0.12\pm0.20$ & $0.19\pm0.18$ & $0.13\pm0.13$ \\
UT200912 & 292.4 & 50.3 & $-3.68\pm5.02$ & $-2.42\pm6.16$ & $0.89\pm1.75$ & $-5.21\pm8.14$ & No & $-0.00\pm0.10$ & $0.01\pm0.14$ & $0.01\pm0.08$ \\
UT191026 & 308.2 & 44.2 & $4.07\pm3.78$ & $-2.24\pm6.00$ & $-0.60\pm1.75$ & $1.23\pm7.31$ & No & $0.15\pm0.11$ & $0.03\pm0.15$ & $0.07\pm0.06$ \\
UT201230g & 320.4 & 47.2 & $-7.25\pm2.23$ & $5.07\pm2.72$ & $0.83\pm0.83$ & $-1.35\pm3.62$ & No & $-0.07\pm0.06$ & $\mathbf{0.14\pm0.05}$ & $0.02\pm0.04$ \\
UT201229a & 348.6 & 47.2 & $-2.41\pm8.91$ & $-20.80\pm15.85$ & $3.28\pm3.01$ & $-19.93\pm18.43$ & No & $0.09\pm0.24$ & $-0.24\pm0.29$ & $0.14\pm0.15$ \\
UT210105 & 351.5 & 47.2 & $1.06\pm2.48$ & $2.50\pm3.52$ & $-0.82\pm1.07$ & $2.74\pm4.44$ & No & $0.06\pm0.05$ & $0.07\pm0.08$ & $-0.01\pm0.04$ \\
UT211106g & 355.4 & 53.8 & $0.32\pm0.56$ & $1.13\pm0.64$ & $0.17\pm0.34$ & $1.62\pm0.91$ & No & $0.02\pm0.01$ & $0.04\pm0.02$ & $0.01\pm0.01$ \\
UT201212 & 356.4 & 47.5 & $7.79\pm4.94$ & $11.04\pm6.14$ & $-0.93\pm2.58$ & $\mathbf{17.90\pm8.29}$ & Yes & $\mathbf{0.21\pm0.10}$ & $0.20\pm0.14$ & $-0.00\pm0.12$ \\
GAvg SQ & 3.8 & 47.0 & $1.48\pm1.58$ & $0.12\pm2.43$ & $-0.65\pm0.72$ & $0.95\pm2.99$ & No & $0.05\pm0.03$ & $0.02\pm0.06$ & $-0.02\pm0.03$ \\
GAvg LQ & 79.1 & 47.2 & $1.43\pm1.74$ & $-0.21\pm2.75$ & $-1.32\pm0.80$ & $-0.10\pm3.35$ & No & $0.05\pm0.04$ & $0.02\pm0.07$ & $-0.02\pm0.03$ \\
GAvg LH & 89.9 & 47.6 & $1.42\pm1.11$ & $-0.48\pm1.70$ & $-1.07\pm0.51$ & $-0.13\pm2.09$ & No & $0.04\pm0.02$ & $0.01\pm0.04$ & $-0.03\pm0.02$ \\
GAvg All & 176.8 & 48.0 & $1.58\pm0.81$ & $-0.31\pm1.19$ & $-0.46\pm0.35$ & $0.82\pm1.49$ & No & $0.03\pm0.02$ & $0.00\pm0.03$ & $-0.01\pm0.01$ \\
GAvg AQ & 184.1 & 50.0 & $1.60\pm1.72$ & $0.12\pm2.52$ & $-0.23\pm0.77$ & $1.48\pm3.15$ & No & $0.03\pm0.04$ & $0.04\pm0.06$ & $-0.01\pm0.03$ \\
GAvg TH & 274.6 & 48.3 & $1.78\pm1.18$ & $-0.08\pm1.70$ & $0.14\pm0.50$ & $1.84\pm2.13$ & No & $0.04\pm0.02$ & $0.01\pm0.04$ & $0.00\pm0.02$ \\
GAvg TQ & 280.7 & 48.1 & $2.06\pm1.57$ & $-1.15\pm2.30$ & $0.13\pm0.65$ & $1.03\pm2.86$ & No & $0.04\pm0.03$ & $-0.00\pm0.06$ & $0.01\pm0.03$
\enddata
\tablecomments{\footnotesize The measured CO$_2$ ice integrated band areas for each spectrum. All errors are 1$\sigma$ errors, and measurements that are positive and $\geq 2\sigma$ from zero are printed in bold. None of the CO$_2$ measurements reached $\geq 3\sigma$ significance. We did not measure the CO$_2$ ice triplet for spectra acquired with the CGS4 instrument or the SpeX PRISM configuration. We use the same UTYYMMDD date code for each spectrum as in the source dataset, including additional letter suffixes: `a' and `b' refer to two separate Miranda spectra observed on the same UT date that had a gap of several hours between acquisition. The `g' suffix indicates spectra obtained with GNIRS. The last several rows of the table are our `TSpec grand average' spectra, which are combined spectra that include all TripleSpec exposures in which the central longitude lies in a specific quadrant (Q) or hemisphere (H). The quadrants and hemispheres are designated with initials: L, leading; T, trailing; A, anti-Uranus; S, sub-Uranus.}
\end{deluxetable}

\begin{deluxetable}{cccccccccccc}
\tabletypesize{\footnotesize}
\tablecaption{2.13-\um and 2.2-\um band measurements\label{tab:nh3bands}}
\tablehead{\colhead{UT Date} & \colhead{Instrument} & \colhead{Long.} & \colhead{Lat.} & \multicolumn{3}{c}{2.13-\um band} & \multicolumn{4}{c}{2.2-\um band} \\ 
\cline{5-8} \cline{9-11}
\colhead{} & \colhead{} & \colhead{} & \colhead{} & \colhead{Band area} & \colhead{Band depth} & \colhead{Both} & \colhead{Band area} & \colhead{Band depth} & \colhead{Both} & \colhead{Center}\\ 
\colhead{} & \colhead{} & \colhead{(\degr E)} & \colhead{(\degr N)} & \colhead{(10$^{-4}\mu$m)} & \colhead{(\%)} & \colhead{$>2\sigma$?} & \colhead{(10$^{-4}\mu$m)} & \colhead{(\%)} & \colhead{$>2\sigma$?} & \colhead{($\mu$m)}}
\startdata
UT210115 & TSpec & 11.8 & 47.2 & $1.50\pm1.51$ & $1.3\pm4.8$ & No & $\mathbf{11.57\pm3.56}$* & $\mathbf{6.3\pm2.4}$ & Yes & $2.223\pm0.016$ \\
UT201229b & TSpec & 29.7 & 47.2 & $-1.29\pm2.38$ & $3.2\pm7.6$ & No & $7.28\pm5.62$ & $\mathbf{9.3\pm3.8}$ & No & \nodata \\
UT000907 & SpeX SXD & 36.3 & -35.4 & $-0.58\pm0.69$ & $1.4\pm1.1$ & No & $0.51\pm1.45$ & $\mathbf{1.9\pm0.9}$ & No & \nodata \\
UT191022a & TSpec & 37.7 & 44.3 & $-4.33\pm3.99$ & $13.9\pm10.5$ & No & $4.74\pm9.17$ & $10.5\pm5.8$ & No & \nodata \\
UT210101 & TSpec & 51.8 & 47.2 & $0.29\pm0.84$ & $3.0\pm2.3$ & No & $3.58\pm1.81$ & $\mathbf{3.0\pm1.2}$ & No & \nodata \\
UT201104 & TSpec & 68.7 & 48.7 & $0.66\pm1.00$ & $\mathbf{6.7\pm2.8}$ & No & $1.95\pm2.44$ & $3.2\pm1.7$ & No & \nodata \\
UT990607 & CGS4 & 75.7 & -36.5 & $-5.41\pm4.65$ & $-6.4\pm5.8$ & No & $\mathbf{30.75\pm9.01}$* & $9.2\pm4.8$ & No & \nodata \\
UT170930 & SpeX PRISM & 80.2 & 36.7 & $0.47\pm1.40$ & $2.5\pm2.9$ & No & $5.91\pm3.84$ & $2.8\pm2.7$ & No & \nodata \\
UT191104 & TSpec & 80.5 & 43.8 & $0.61\pm1.05$ & $3.9\pm3.5$ & No & $3.46\pm2.42$ & $3.2\pm1.6$ & No & \nodata \\
UT150912 & SpeX PRISM & 92.1 & 30.4 & $0.19\pm2.74$ & $4.8\pm6.8$ & No & $4.19\pm6.16$ & $7.8\pm4.5$ & No & \nodata \\
UT200907 & TSpec & 96.1 & 50.4 & $-1.23\pm1.25$ & $5.0\pm3.7$ & No & $\mathbf{8.13\pm2.85}$ & $\mathbf{5.3\pm1.7}$* & Yes & $2.214\pm0.013$ \\
UT171010 & SpeX PRISM & 107.1 & 36.1 & $-1.18\pm1.38$ & $-1.4\pm2.5$ & No & $-6.74\pm3.36$ & $1.1\pm1.9$ & No & \nodata \\
UT201008g & GNIRS & 109.4 & 49.6 & $0.19\pm0.50$ & $1.5\pm1.6$ & No & $\mathbf{5.66\pm1.16}$* & $\mathbf{2.8\pm0.5}$* & Yes* & $2.218\pm0.016$ \\
UT120926 & SpeX SXD & 152.6 & 21.4 & $0.61\pm3.39$ & $5.2\pm9.0$ & No & $9.73\pm10.92$ & $12.2\pm6.1$ & No & \nodata \\
UT201007 & TSpec & 174.2 & 49.7 & $\mathbf{2.41\pm0.90}$ & $5.6\pm3.9$ & No & $0.49\pm2.17$ & $2.8\pm1.6$ & No & \nodata \\
UT200913 & TSpec & 184.9 & 50.3 & $-0.72\pm1.39$ & $1.5\pm4.2$ & No & $-0.33\pm3.28$ & $4.6\pm2.6$ & No & \nodata \\
UT200930 & TSpec & 192.8 & 49.9 & $-0.07\pm0.91$ & $4.4\pm3.1$ & No & $2.46\pm2.14$ & $\mathbf{3.3\pm1.3}$ & No & \nodata \\
UT211121g & GNIRS & 203.7 & 53.3 & $-0.09\pm0.48$ & $0.4\pm1.5$ & No & $\mathbf{5.90\pm1.11}$* & $\mathbf{2.7\pm0.5}$* & Yes* & $2.219\pm0.017$ \\
UT170925 & SpeX SXD & 232.5 & 36.7 & $1.40\pm1.92$ & $6.2\pm8.3$ & No & $-15.49\pm4.68$ & $3.9\pm2.8$ & No & \nodata \\
UT150911 & SpeX PRISM & 236.2 & 30.4 & $-5.87\pm3.53$ & $-2.9\pm7.4$ & No & $-3.87\pm9.99$ & $13.3\pm6.7$ & No & \nodata \\
UT120925 & SpeX SXD & 256.1 & 21.4 & $-1.34\pm4.18$ & $3.0\pm12.3$ & No & $5.57\pm10.71$ & $\mathbf{11.5\pm5.2}$ & No & \nodata \\
UT201013 & TSpec & 262.8 & 49.5 & $0.06\pm0.78$ & $0.3\pm2.6$ & No & $\mathbf{8.27\pm1.88}$* & $\mathbf{4.3\pm1.6}$ & Yes & $2.211\pm0.018$ \\
UT201206 & TSpec & 274.6 & 47.7 & $-0.19\pm0.87$ & $2.2\pm2.7$ & No & $0.41\pm1.96$ & $\mathbf{2.7\pm1.3}$ & No & \nodata \\
UT141130 & SpeX SXD & 279.4 & 24.7 & $1.52\pm1.84$ & $5.0\pm8.1$ & No & $\mathbf{10.49\pm4.88}$ & $\mathbf{8.4\pm2.9}$ & Yes & $2.219\pm0.016$ \\
UT150917 & SpeX PRISM & 280.3 & 30.2 & $-3.60\pm1.70$ & $-1.8\pm3.9$ & No & $-6.52\pm4.20$ & $3.2\pm2.7$ & No & \nodata \\
UT191013a & TSpec & 285.7 & 44.6 & $-1.37\pm2.37$ & $6.5\pm6.6$ & No & $-2.41\pm5.45$ & $\mathbf{6.9\pm3.4}$ & No & \nodata \\
UT200912 & TSpec & 292.4 & 50.3 & $-2.13\pm1.21$ & $0.7\pm3.8$ & No & $4.78\pm2.91$ & $4.2\pm2.1$ & No & \nodata \\
UT191026 & TSpec & 308.2 & 44.2 & $1.87\pm1.45$ & $2.9\pm5.6$ & No & $\mathbf{11.79\pm3.25}$* & $\mathbf{6.6\pm2.4}$ & Yes & $2.210\pm0.015$ \\
UT201230g & GNIRS & 320.4 & 47.2 & $0.28\pm0.85$ & $2.9\pm3.0$ & No & $3.22\pm1.99$ & $\mathbf{3.8\pm1.0}$* & No & \nodata \\
UT201229a & TSpec & 348.6 & 47.2 & $-0.72\pm2.44$ & $6.4\pm8.1$ & No & $3.49\pm5.75$ & $7.9\pm4.5$ & No & \nodata \\
UT210105 & TSpec & 351.5 & 47.2 & $1.33\pm0.78$ & $3.0\pm2.0$ & No & $3.39\pm1.79$ & $\mathbf{2.7\pm1.1}$ & No & \nodata \\
UT211106g & GNIRS & 355.4 & 53.8 & $0.32\pm0.30$ & $0.4\pm1.0$ & No & $\mathbf{6.35\pm0.70}$* & $\mathbf{2.1\pm0.3}$* & Yes* & $2.211\pm0.010$ \\
UT201212 & TSpec & 356.4 & 47.5 & $1.30\pm1.86$ & $5.3\pm6.5$ & No & $-4.64\pm4.26$ & $3.1\pm2.7$ & No & \nodata \\
GAvg SQ & TSpec & 3.8 & 47.0 & $\mathbf{1.12\pm0.53}$ & $2.1\pm1.4$ & No & $\mathbf{4.46\pm1.23}$* & $\mathbf{2.5\pm0.9}$ & Yes & $2.209\pm0.011$ \\
GAvg LQ & TSpec & 79.1 & 47.2 & $0.00\pm0.59$ & $2.1\pm1.7$ & No & $\mathbf{4.28\pm1.33}$* & $\mathbf{2.4\pm0.9}$ & Yes & $2.215\pm0.015$ \\
GAvg LH & TSpec & 89.9 & 47.6 & $0.73\pm0.38$ & $1.8\pm1.1$ & No & $\mathbf{3.53\pm0.86}$* & $\mathbf{2.3\pm0.7}$* & Yes* & $2.203\pm0.014$ \\
GAvg All & TSpec & 176.8 & 48.0 & $0.44\pm0.27$ & $1.3\pm0.8$ & No & $\mathbf{3.89\pm0.62}$* & $\mathbf{2.1\pm0.4}$* & Yes* & $2.204\pm0.013$ \\
GAvg AQ & TSpec & 184.1 & 50.0 & $0.88\pm0.58$ & $2.4\pm1.9$ & No & $1.13\pm1.40$ & $\mathbf{2.4\pm1.0}$ & No & \nodata \\
GAvg TH & TSpec & 274.6 & 48.3 & $0.14\pm0.38$ & $1.0\pm1.2$ & No & $\mathbf{4.08\pm0.89}$* & $\mathbf{2.1\pm0.6}$* & Yes* & $2.215\pm0.013$ \\
GAvg TQ & TSpec & 280.7 & 48.1 & $-0.14\pm0.51$ & $0.4\pm1.6$ & No & $\mathbf{5.12\pm1.16}$* & $\mathbf{2.5\pm1.0}$ & Yes & $2.208\pm0.013$
\enddata
\tablecomments{\footnotesize The measured integrated band areas and band depths for the 2.13-\um and 2.2-\um bands in each spectrum. Central wavelengths for the 2.2-\um band are only printed if both the area and depth measurements are $\geq 2\sigma$ from zero. Measurements which are $\geq 3\sigma$ from zero have an asterisk. All errors are 1$\sigma$ errors. Other notation follows that of Table \ref{tab:CO2bands}.}
\end{deluxetable}

\begin{deluxetable}{ccccccccc}
\tabletypesize{\footnotesize}
\tablecaption{Mean band measurements\label{tab:meanbands}}
\tablehead{\colhead{Quadrant} & \colhead{Longitude} & \colhead{Total CO$_2$ area} & \colhead{2.13-\um area} & \colhead{2.13-\um depth} & \colhead{2.2-\um area} & \colhead{2.2-\um depth} & \colhead{2.13-\um center} & \colhead{2.2-\um center} \\
\colhead{} & \colhead{(range, \degr E)} & \colhead{(10$^{-4}\mu$m)} & \colhead{(10$^{-4}\mu$m)} & \colhead{(\%)} & \colhead{(10$^{-4}\mu$m)} & \colhead{(\%)} & \colhead{($\mu$m)} & \colhead{($\mu$m)}}
\startdata
 all & 0 -- 360 & $-0.07\pm2.12$ & $-0.46\pm0.49$ & $\mathbf{2.9\pm1.2}$ & $\mathbf{3.76\pm1.58}$ & $\mathbf{5.4\pm0.8}$* & $2.130\pm0.001$ & $2.218\pm0.003$ \\
 LH & 1 -- 180 & $-2.26\pm3.16$ & $-0.47\pm0.78$ & $3.3\pm1.8$ & $\mathbf{6.08\pm2.50}$ & $\mathbf{5.4\pm1.2}$* & $2.130\pm0.001$ & $2.219\pm0.004$ \\
 TH & 181 -- 360 & $1.44\pm2.84$ & $-0.44\pm0.64$ & $2.6\pm1.5$ & $1.82\pm1.93$ & $\mathbf{5.3\pm1.0}$* & $2.130\pm0.001$ & $2.218\pm0.004$ \\
 LQ & 45 -- 135 & $0.02\pm2.07$ & $-0.60\pm0.94$ & $2.2\pm1.9$ & $6.32\pm3.65$ & $\mathbf{4.3\pm1.3}$* & $2.130\pm0.001$ & $2.218\pm0.005$ \\
 TQ & 225 -- 315 & $3.04\pm3.33$ & $-0.96\pm1.05$ & $2.2\pm2.4$ & $1.30\pm3.26$ & $\mathbf{6.5\pm1.6}$* & $2.130\pm0.001$ & $2.216\pm0.006$ \\
 AQ & 135 -- 225 & $-0.40\pm4.49$ & $0.43\pm0.95$ & $3.4\pm2.5$ & $3.65\pm3.01$ & $\mathbf{5.1\pm2.3}$ & $2.130\pm0.001$ & $2.219\pm0.008$ \\
 SQ & 315 -- 45 & $-2.69\pm6.07$ & $-0.24\pm0.90$ & $4.2\pm2.4$ & $3.99\pm2.14$ & $\mathbf{5.3\pm1.5}$* & $2.130\pm0.001$ & $2.221\pm0.006$ \\
\enddata
\tablecomments{\footnotesize Mean band area and depth measurements, averaged over quadrants and hemispheres. All errors are 1$\sigma$ errors. Measurements that are greater than zero with $\geq 2\sigma$ significance are printed in bold, and $\geq 3\sigma$ with an asterisk. The quadrants (Q) and hemispheres (H) are designated with initials: L, leading; T, trailing; A, anti-Uranus; S, sub-Uranus. For example, LH indicates the leading hemisphere average of all spectra with longitudes between 0 -- 180\degr E.}
\end{deluxetable}

\begin{deluxetable}{ccccccc}
\tabletypesize{\footnotesize}
\tablecaption{Quadrant/hemisphere ratios\label{tab:ratiobands}}
\tablehead{\colhead{Dataset} & \colhead{Ratio} & \colhead{Total CO$_2$ area} & \colhead{2.13-\um area} & \colhead{2.13-\um depth} & \colhead{2.2-\um area} & \colhead{2.2-\um depth}}
\startdata
All spectra & LQ/TQ & $0.01\pm0.68$ & $0.62\pm1.19$ & $0.98\pm1.35$ & $4.86\pm12.50$ & $0.66\pm0.25$ \\
 & LH/TH & $-1.57\pm3.79$ & $1.06\pm2.33$ & $1.29\pm1.03$ & $3.33\pm3.78$ & $1.03\pm0.31$ \\
 & AQ/SQ & $0.15\pm1.70$ & $-1.77\pm7.62$ & $0.81\pm0.75$ & $0.91\pm0.90$ & $0.97\pm0.51$
\enddata
\tablecomments{\footnotesize Ratios of the band parameters between opposing quadrants and hemispheres. All errors are 1$\sigma$ errors. }
\end{deluxetable}

\begin{deluxetable}{ccccccc}
\tabletypesize{\footnotesize}
\tablecaption{Sinusoidal models and F-test\label{tab:Ftable}}
\tablehead{\colhead{Dataset} & \colhead{Measurement} & \colhead{N} & \colhead{Long. Max} & \colhead{F-value} & \colhead{p-value} & \colhead{Reject?}}
\startdata
All spectra & Total CO$_2$ band area & 27 & \nodata & 0.198 & 0.82177 & No \\
 & 2.13-$\mu m$ area & 33 & \nodata & 0.149 & 0.86231 & No \\
 & 2.13-$\mu m$ depth & 33 & \nodata & 0.444 & 0.64557 & No \\
 & 2.2-$\mu m$ area & 33 & \nodata & 0.342 & 0.71298 & No \\
 & 2.2-$\mu m$ depth & 33 & \nodata & 0.902 & 0.41661 & No \\
 & 2.2-$\mu m$ centers & 33 & \nodata & 0.064 & 0.93848 & No \\
Mean bands & Total CO$_2$ band area & 4 & \nodata & 0.425 & 0.73538 & No \\
 & 2.13-$\mu m$ area & 4 & \nodata & 0.196 & 0.84733 & No \\
 & 2.13-$\mu m$ depth & 4 & \nodata & 0.056 & 0.94810 & No \\
 & 2.2-$\mu m$ area & 4 & 89.8 & 844.862 & 0.02432 & Yes \\
 & 2.2-$\mu m$ depth & 4 & \nodata & 105.062 & 0.06882 & No \\
 & 2.2-$\mu m$ centers & 4 & \nodata & 0.202 & 0.84397 & No \\
\enddata
\tablecomments{\footnotesize Sinusoidal model fits versus longitude to the band parameters for two sets of measurements: the individual band parameters for each spectrum, and the mean band parameters (averaged over the four longitude quadrants). The total CO$_2$ band area has fewer measurements because we did not measure CO$_2$ ice in the CGS4 or SpeX PRISM spectra. The degrees of freedom for the sinusoidal model fit is N - 3. We only print the longitude of maximum band strength for the measurements where we rejected the null hypothesis ($p < 0.05$). }
\end{deluxetable}

\subsection{\texorpdfstring{CO$_2$}{CO2} ice triplet}

The results of our CO$_2$ triplet band measurements are shown in Table \ref{tab:CO2bands} and Figures \ref{fig:bandareas} and \ref{fig:meanbandareas}. Even the highest S/N Miranda spectra do not show adequate evidence for the presence of the CO$_2$ ice triplet around 2 \micron. We measured the areas and depths of each individual CO$_2$ band, but $\geq$2$\sigma$ significance in any of the band area or depth measurements were only measured in three spectra. To assist in the detection of very faint features, we also calculated the sum of all three CO$_2$ ice band areas. However, only a single spectrum (UT201212) passes the 2$\sigma$ threshold for the total CO$_2$ band area. Visual inspection of that spectrum and other spectra in which only a single band was detected indicates that these `detections' are simply spectral noise, probably due to imperfect telluric correction, especially in the 1.966-\um band affected by atmospheric water vapor. The quadrant- and hemisphere-averaged means of the total CO$_2$ band area (Table \ref{tab:meanbands}) only reinforces our non-detection. We conclude that discrete deposits of CO$_2$ ice are not present on Miranda in amounts detectable by our observations.

\subsection{\texorpdfstring{2.13-\um}{2.13-um} feature}

We also measured the band area and depth of the 2.13-\um feature in spectra of Miranda, as this absorption band could potentially be associated with a molecular mixture of CO$_2$ and H$_2$O ice \citep{Bernstein2005}. Weak absorption features near this wavelength have been noted on both Ariel and Umbriel and suggested to be potential evidence of CO$_2$:H$_2$O ice molecular mixtures as part of the radiolytic CO$_2$ production cycle \citep{Cart2022,Cart2023}. Our band parameter measurements for the 2.13-\um band are shown in Table \ref{tab:nh3bands} and Figures \ref{fig:bandareas} and \ref{fig:meanbandareas}. Two spectra show $\geq 2\sigma$ significance in only the 2.13-\um area (UT201007 and the TripleSpec sub-Uranus quadrant grand average) or only the 2.13-\um depth (UT201104), but none show $\geq 2\sigma$ significance in both. Our mean measurements do show $\geq 2\sigma$ significance in the 2.13-\um depth when taking the mean of all spectra, regardless of longitude. However, the lack of consistency in the significance of the band area and depth measurements leads us to conclude that there is insufficient evidence for a 2.13-\um band on Miranda.

\subsection{\texorpdfstring{2.2-\um}{2.2-um} feature}
The 2.2-\um bands in spectra of Miranda are weak, hindering clear identifications of the compounds that might be responsible. Furthermore, absorption bands in the 2.2-\um region can vary in central wavelength, with absorptions previously reported on other bodies at wavelengths ranging from 2.18 \um to 2.25 \micron. However, we measured non-zero band areas and depths in many of our spectra: 8 out of the 33 spectra from individual nights have $\geq 2\sigma$ measurements of both 2.2-\um band areas and 2.2-\um band depths, although a 2.2-\um feature is not always visually apparent. For the GNIRS spectra, we measured $\geq 3\sigma$ significance in both 2.2-\um area and depth for three out of the four spectra. We measured $\geq 3\sigma$ significance for both area and depth in three of the seven TripleSpec grand average spectra, and $\geq 2\sigma$ significance for six of the seven.  The trailing hemisphere GNIRS spectrum and TripleSpec grand average for the anti-Uranus quadrant were both only $2\sigma$ significant in depth and not area. The trailing hemisphere GNIRS spectrum was lower S/N than the other three GNIRS spectra (Figure \ref{fig:CO2NH3vstack}).

Furthermore, when taking the means of the individual band measurements (not including the grand averages), we measured $\geq 2\sigma$ 2.2-\um band depths for all of the quadrant and hemisphere-averaged means, and the mean of all spectra and the mean of the leading-hemisphere spectra were measured to $\geq 2\sigma$ significance for both the 2.2-\um band area and depth. We calculated an average band depth (across all spectra) of 5.4$\pm$0.8\%. As discussed in the next section, we do not find convincing evidence for sub-observer longitudinal trends (i.e. hemispherical asymmetries) in the 2.2-\um band.

The band center measurements for the 2.2-\um band in the individual spectra generally clustered between 2.21 and 2.22 \micron. The mean band center for all spectra was 2.218$\pm$0.003 \um (Table \ref{tab:meanbands}). For individual spectra in which the band was detected at $>2\sigma$ significance, the minimum band center wavelength of 2.203$\pm$0.014 \um was measured for the TripleSpec leading hemisphere grand average. The maximum wavelength was measured at 2.223$\pm$0.016 \micron, for the TripleSpec spectrum from UT210115. This average central wavelength of 2.218 \um is at slightly longer wavelengths than NH$_3$-hydrates and H$_2$O:NH$_3$ ice mixtures, which tend to range between 2.209 and 2.216 \micron. Pure cubic NH$_3$ ice exhibits a band at 2.241 \um \citep{Moore2007}. However, the range of band centers in our spectra is not definitive, as we measured band centers based on a single band spanning 2.186 -- 2.251 \micron. Our band centers would be skewed towards longer wavelengths by the presence of crystalline NH$_3$ ice at 2.24 \micron, and would also be affected by the presence of a `spike' in the data at 2.207 \micron, in the middle of any band at 2.20 -- 2.21 \micron, that we attribute to residuals from differences in metal abundances in our telluric standard stars and the solar spectrum (\S \ref{sec:contam}). We discuss possibilities for candidate materials, including amorphous NH$_3$ ice, other ices, and NH$_4$-bearing minerals, in sections \S \ref{sec:modelresults}, \ref{ssec:nh3discussion}, and \ref{ssec:nh4discussion}.

Finally, what of the \citet{Bauer2002} detection of the 2.2-\um feature? For this spectrum (UT990607), we measured a statistically significant 2.2-\um band area, but the band depth did not reach 2$\sigma$ significance. Unfortunately, the axial tilt and seasonal illumination conditions of the Uranian system also makes direct reproduction (reobservation) of the Bauer et al. study impossible in the next several decades. Their spectrum was observed at southern latitudes (36.5\degr S), while all but one of the spectra in our dataset were observed on the northern hemisphere. It is worth noting that the Bauer et al. spectrum showing absorption at 2.2-\um was observed when Arden Corona was near disk center (Figure \ref{fig:MirGeology}), supporting the notion that the 2.2-\um band may be associated with geologically young terrain. Our dataset also has a single SpeX spectrum of Miranda (UT000907) observed at a similar southern latitude (35.4\degr S), but at a different sub-observer longitude that includes more heavily cratered, older terrain. This SpeX spectrum does not show strong evidence of a 2.2-\um band.

\subsection{Sub-observer longitudinal trends}

When considering the quadrant/hemisphere ratios of the band parameters, we did not find statistically significant ($\geq 2\sigma$) departures from a ratio of unity for any of the measurements or quadrant/hemisphere pairs (Table \ref{tab:ratiobands}). Our non-detection of the CO$_2$ ice bands and 2.13-\um band precludes discussion of longitudinal trends for these bands, although the ratio calculations are included in the tables for completeness. 

When fitting a sinusoidal model to the 2.2-\um band measurements from the 33 individual spectra, none of the sinusoidal models could be considered a statistically significant better fit than a constant weighted average. We found that only the model fit to the 2.2-\um area measurements for the quadrant-averaged mean band parameters was statistically significant (Table \ref{tab:Ftable}). We do not find the sinusoidal fit for the quadrant-averaged mean 2.2-\um band \textit{depth} to be significant (although it is close, at $p$ $\sim$ 0.07). We are inclined to trust the sinusoidal fit to 33 data points instead of four data points, even though the four data points are weighted averages of the 33 individual measurements. Finally, we do not find statistically significant deviations from unity for any of the calculated quadrant or hemisphere ratios of band parameters.  

When considering the results from the quadrant/hemisphere ratios and F-tests of the sinusoidal models, we conclude that there is minimal evidence for large-scale longitudinal trends or hemispherical asymmetries in the presence or strength of the 2.2-\um band.

\section{Spectral modeling \label{sec:modelresults}}
\begin{figure}[ht!]
\centering
\makebox[\textwidth][c]{\includegraphics[width=\textwidth]{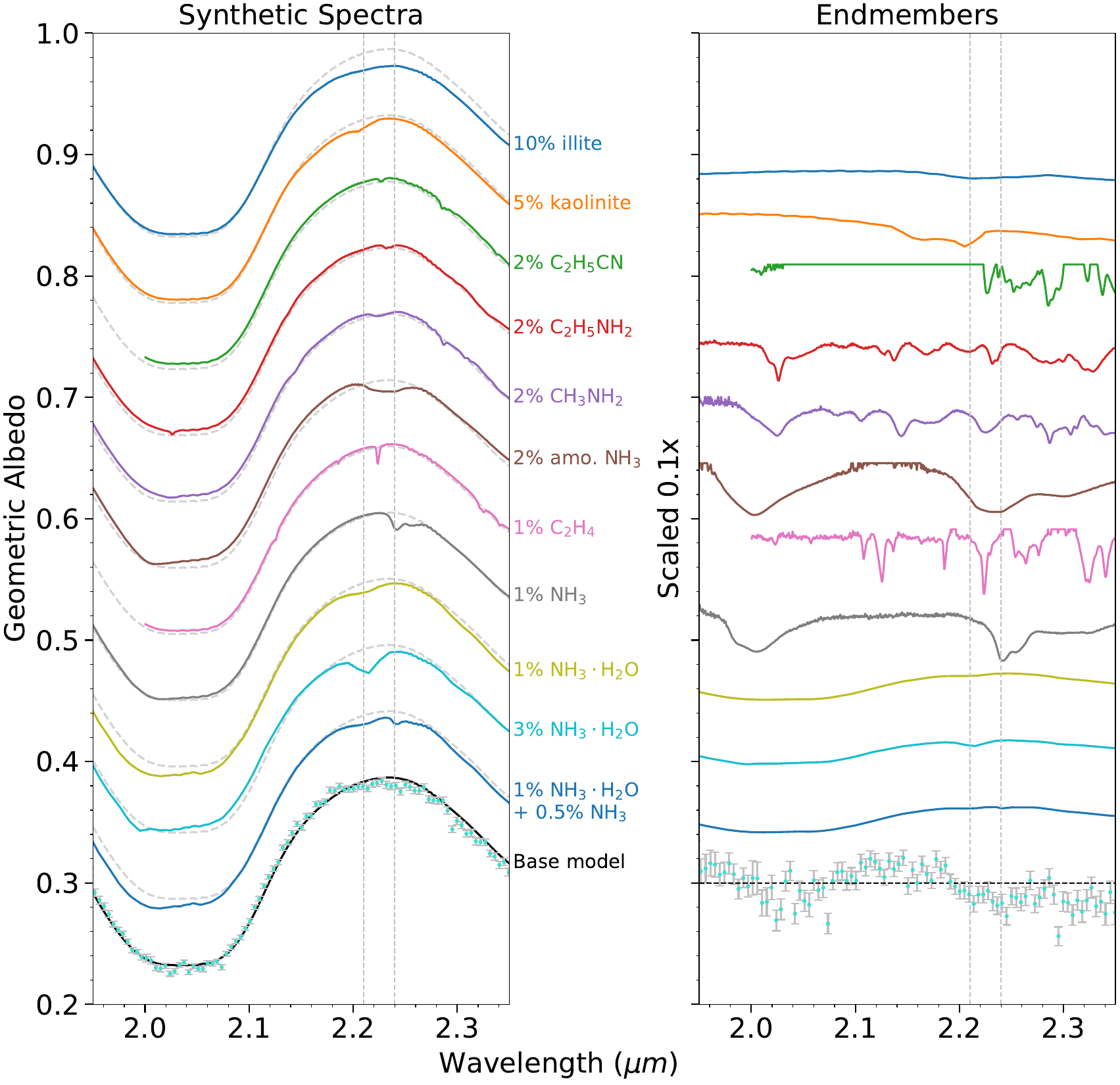}}
\vspace{-15pt}
\caption{\footnotesize (Left panel): Synthetic spectra generated with a Hapke-Mie modeling program to investigate potential absorption features in the 2.2-\um region, compared to the UT201008 GNIRS spectrum of Miranda's leading hemisphere (bottom, binned by 10 pixels). The synthetic spectra are described in Table \ref{tab:synspectra}. The left panel includes comparison spectra in which a small percentage of the 27.5-\um H$_2$O ice in the base model is replaced with an additional component of different composition, with a grain size of 10 \micron. Vertical lines are placed at 2.21 and 2.24 \micron. (Right panel): as the left panel, but using synthetic spectra of the corresponding pure (100\%) ices at 10 \um grain size, scaled by a factor of 0.1. The observed GNIRS spectrum is divided by the base model. }
\vspace{-5pt}
\label{fig:22umhapkeices}
\end{figure}

\begin{deluxetable}{ccccccc}
\tabletypesize{\footnotesize}
\tablecaption{Synthetic spectra for Figure \ref{fig:22umhapkeices}\label{tab:synspectra}}
\tablehead{\colhead{Short name} & \colhead{Component 1} & \colhead{Component 2} & \colhead{Component 3} & \colhead{Component 4} & \colhead{$\chi_{\nu}^{2}$} & \colhead{Source}}
\startdata
Base model & 95.25\% H$_2$O 27.5$\mu m$ & 4.5\% AC 5$\mu m$ & 0.25\% H$_2$O 0.3$\mu m$ & \nodata & 1.050 & see caption \\
10\% illite & 85.25\% H$_2$O 27.5$\mu m$ & 4.5\% AC 5$\mu m$ & 0.25\% H$_2$O 0.3$\mu m$ & 10\% illite 10$\mu m$ & 2.762 & \citet{Clark1990Minerals} \\
5\% kaolinite & 90.25\% H$_2$O 27.5$\mu m$ & 4.5\% AC 5$\mu m$ & 0.25\% H$_2$O 0.3$\mu m$ & 5\% kaolinite 10$\mu m$ & 1.202 & \citet{Clark1990Minerals} \\
2\% propionitrile & 93.25\% H$_2$O 27.5$\mu m$ & 4.5\% AC 5$\mu m$ & 0.25\% H$_2$O 0.3$\mu m$ & 2\% C$_2$H$_5$CN 10$\mu m$ & 1.351 & \citet{Moore2010} \\
2\% ethylamine & 93.25\% H$_2$O 27.5$\mu m$ & 4.5\% AC 5$\mu m$ & 0.25\% H$_2$O 0.3$\mu m$ & 2\% C$_2$H$_5$NH$_2$ 10$\mu m$ & 1.252 & \citet{Hudson2022} \\
2\% methylamine & 93.25\% H$_2$O 27.5$\mu m$ & 4.5\% AC 5$\mu m$ & 0.25\% H$_2$O 0.3$\mu m$ & 2\% CH$_3$NH$_2$ 10$\mu m$ & 1.236 & \citet{Hudson2022} \\
2\% amo. NH$_3$ & 93.25\% H$_2$O 27.5$\mu m$ & 4.5\% AC 5$\mu m$ & 0.25\% H$_2$O 0.3$\mu m$ & 2\% amorph. NH$_3$ 10$\mu m$ & 0.938 & \citet{Roser2021} \\
1\% ethylene & 94.25\% H$_2$O 27.5$\mu m$ & 4.5\% AC 5$\mu m$ & 0.25\% H$_2$O 0.3$\mu m$ & 1\% C$_2$H$_4$ 10$\mu m$ & 1.212 & \citet{Hudson2014} \\
1\% NH$_3$ & 94.25\% H$_2$O 27.5$\mu m$ & 4.5\% AC 5$\mu m$ & 0.25\% H$_2$O 0.3$\mu m$ & 1\% cubic NH$_3$ 10$\mu m$ & 0.908 & \citet{Hudson2022} \\
1\% NH$_3\cdot$H$_2$O & 94.25\% (1\%)NH$_3\cdot$H$_2$O 27.5$\mu m$ & 4.5\% AC 5$\mu m$ & 0.25\% H$_2$O 0.3$\mu m$ & \nodata & 1.051 & see caption \\
3\% NH$_3\cdot$H$_2$O & 94.25\% (3\%)NH$_3\cdot$H$_2$O 27.5$\mu m$ & 4.5\% AC 5$\mu m$ & 0.25\% H$_2$O 0.3$\mu m$ & \nodata & 1.602 & see caption \\
Mixed NH$_3$ & 94.75\% (1\%)NH$_3\cdot$H$_2$O 27.5$\mu m$ & 4.5\% AC 5$\mu m$ & 0.25\% H$_2$O 0.3$\mu m$ & 0.5\% cubic NH$_3$ 10$\mu$m & 0.939 & see caption \\
\enddata
\tablecomments{\footnotesize The components of the synthetic Hapke-Mie spectra plotted in Figure \ref{fig:22umhapkeices}, with their mixing ratios and grain sizes for each component. All spectra are intimate (particulate) mixtures. Components listed as `H$_2$O' use optical constants of 80 K crystalline ice from \citet{Mastrapa2008}, while the component listed as `AC' is amorphous carbon, using the optical constants of sample BE1 from \citet{RouleauMartin1991}. Optical constants for other components are listed in the table. The NH$_3$-hydrate spectra replace the primary H$_2$O ice component with the optical constants of 1\% and 3\% by weight mixtures (respectively) of NH$_3\cdot$H$_2$O. Reflectance spectra of these mixtures were originally measured by \citet{Brown1988}, with optical constants derived by T. Roush (personal communication) and published in \citet{Cruikshank2005}.}
\end{deluxetable}

\begin{figure}[ht!]
\centering
\makebox[\textwidth][c]{\includegraphics[width=\textwidth]{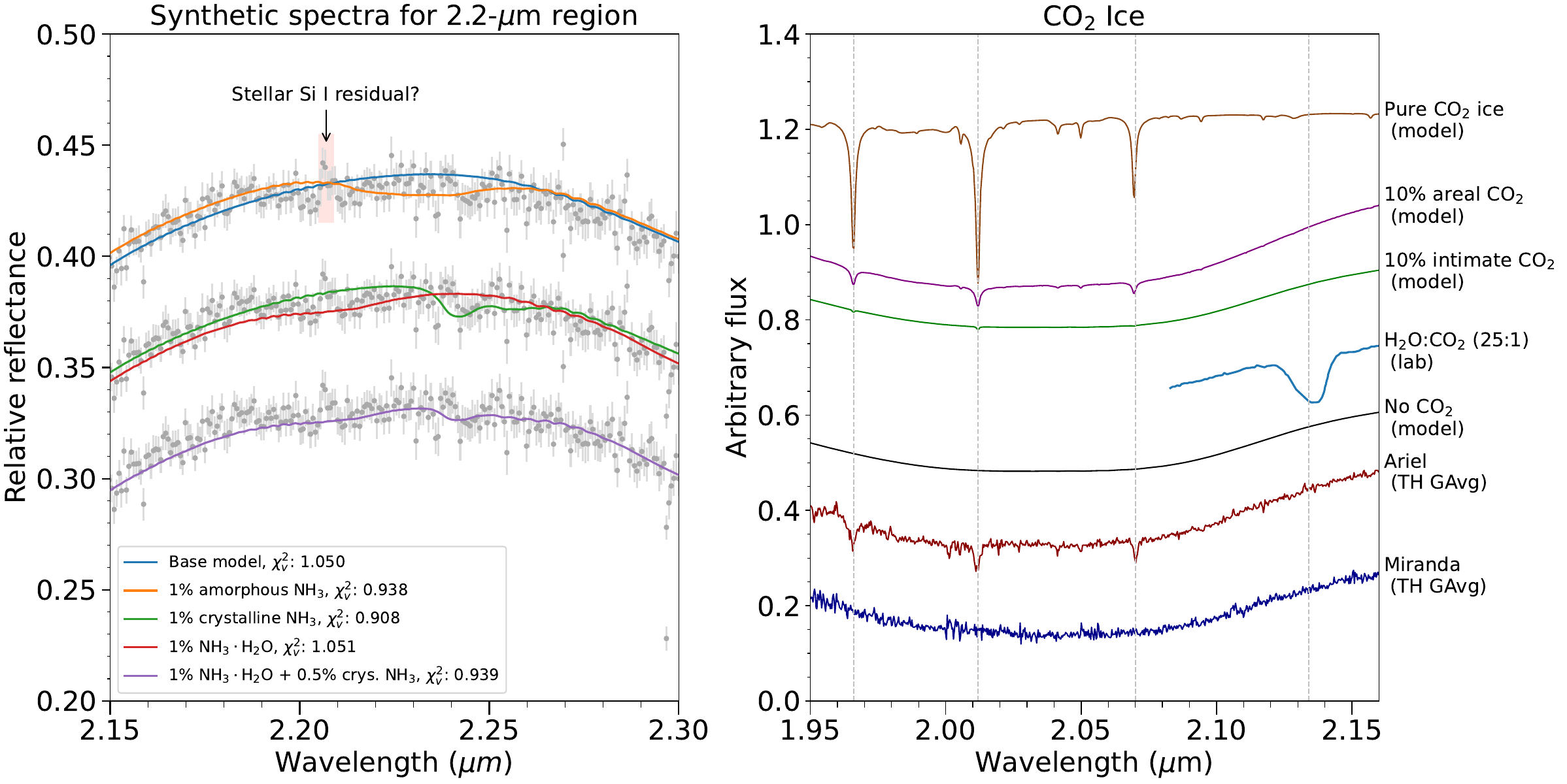}}
\vspace{-15pt}
\caption{\footnotesize (Left panel): A close-up of the most promising NH$_3$-bearing synthetic models from Figure \ref{fig:22umhapkeices}, compared to the UT201008 GNIRS spectrum of Miranda's leading hemisphere. The synthetic spectra are described in Table \ref{tab:synspectra}. Small periodic variations are visible in the synthetic spectra due to resonances from the calculated Mie scattering solutions. (Right panel): A demonstration of the lack of CO$_2$ ice features on Miranda's trailing hemisphere, compared to the trailing hemisphere of Ariel \citep{Cart2022}. We include several Hapke model spectra: pure CO$_2$ ice (10 \um grains), a base model (Table \ref{tab:synspectra}), the base model with 10\% areally mixed CO$_2$ ice (27.5 \um grains), and the base model with 10\% intimately mixed CO$_2$ ice (27.5 \um grains). We also include a laboratory spectrum of a H$_2$O:CO$_2$ ice molecular mixture \citep{Bernstein2005}, which displays a prominent 2.13-\um 2$\nu_3$ overtone band. Note that this forbidden overtone does not appear in pure CO$_2$ ice or the model spectra of H$_2$O:CO$_2$ ice intimate mixtures. }
\vspace{-5pt}
\label{fig:hapke22zoomCO2}
\end{figure}

Synthetic Hapke-Mie spectra of intimate mixtures including some of the same materials as in \citet{Cart2023} are plotted in Figure \ref{fig:22umhapkeices} and tabulated in Table \ref{tab:synspectra}. We compared these synthetic spectra to a spectrum of Miranda's leading hemisphere obtained with GNIRS (spectrum UT201008). This spectrum was chosen for being both high S/N and showing weak absorption in the 2.2-\um region. This absorption is composed of a shallow dip between 2.19 -- 2.23 \micron, along with a possible feature at 2.24 \um (Figure \ref{fig:hapke22zoomCO2}). We constructed a base model composed of two grain sizes of H$_2$O ice (27.5 \um and 0.3 \micron), along with an amorphous carbon component to act as a dark, spectrally neutral absorber. 

To demonstrate the lack of CO$_2$ ice absorption features, we calculated spectral models incorporating CO$_2$ ice and plotted them against observed spectra in Figure \ref{fig:hapke22zoomCO2}. Spectral models of pure or intimately-mixed CO$_2$ ice fail to reproduce the 2.13-\um 2$\nu_3$ overtone band seen in laboratory spectra of CO$_2$ ice in a molecular mixture with H$_2$O \citep{Bernstein2005}. A spectrum of Ariel's trailing hemisphere \citep{Cart2022} shows clear signatures of CO$_2$ ice, as previously discussed (\S \ref{ssec:CO2bkg}). In comparison, the Miranda trailing hemisphere grand average spectrum shows no evidence of any absorption features resulting from CO$_2$ ice, either as separate deposits or trapped in the regolith. 

To investigate species that could be contributing to absorption features in the 2.2-\um range, we constructed models in which we replaced small percentages of the H$_2$O ice component in the base model with other materials, including some of the same materials as in \citet{Cart2023} (Figure \ref{fig:22umhapkeices}). From visual inspection, the complex absorption bands of propionitrile (C$_2$H$_5$CN), ethylamine (C$_2$H$_5$NH$_2$) and methylamine (CH$_3$NH$_2$) are somewhat difficult to interpret, but do not match the Miranda spectrum closely. Ethylamine and methylamine have band complexes between 1.65 -- 1.80 \um which are also not apparent in our spectra. The 2.22-\um band of ethylene (C$_2$H$_4$) is too sharp compared to the broad absorptions of the Miranda spectra. The 2.2-\um bands of the aluminum-bearing phyllosilicates illite and kaolinite are subtle, but both of these minerals also have bands at 1.4 \um from bound water absorption. The wavelengths between 1.38 -- 1.43 \um are hard to observe from Earth due to strong atmospheric H$_2$O absorption. Spectra with good telluric correction observed in dry conditions (such as the GNIRS spectra) do not show convincing evidence for the presence of the 1.4-\um band, but it is difficult to rule out either kaolinite or illite given the lower S/N in the region.

We also calculated $\chi_{\nu}^{2}$ values as a comparison between the observed spectrum and the constructed models, using the wavelengths between 2.15 -- 2.29 \um (Table \ref{tab:synspectra}). The synthetic intimate mixture spectra we generated are not unique solutions, but are useful as general guidelines for interpretation of spectral features. Both cubic (crystalline) and amorphous NH$_3$ mixtures have significantly improved $\chi_{\nu}^{2}$ values compared to the base model, while a 1\% mixture of NH$_3\cdot$ H$_2$O fits the data equally as well as the base model ($\chi_{\nu}^{2}$ of 1.051 versus 1.050). Kaolinite, propionitrile, ethylene, ethylamine, methylamine, the 3\% mixture of NH$_3\cdot$ H$_2$O, and illite all have worse $\chi_{\nu}^{2}$ values than the base model. 

While amorphous NH$_3$ ice has one of the most favorable $\chi_{\nu}^{2}$ values, it is unlikely that amorphous NH$_3$ could persist on Miranda's surface, as it transitions to crystalline NH$_3$ when warmed to temperatures between 70 -- 90 K \citep{Moore2007}. Voyager 2 measured a surface brightness temperature for Miranda of about 86 K \citep{Hanel1986}, and thermodynamical modeling predicts peak surface temperatures for the Uranian satellites around 90 K \citep{Sori2017}. Crystalline NH$_3$ ice produces a band at 2.24 \micron, which is at longer wavelengths than the broad 2.2-\um feature in most of our spectra, but does correspond to the weak feature at 2.24 \um seen in this observed GNIRS spectrum (Figure \ref{fig:hapke22zoomCO2}). We also find a promising match between the spectrum in which the major H$_2$O ice component is replaced with a 1\% mixture of NH$_3$-hydrates, which produces a shallow shoulder feature. However, the 3\% NH$_3$-hydrates mixture produces a distinct 2.2-\um absorption band that is too strong compared to the weak absorption in the observed spectrum. We further investigated the possibility of NH$_3$ species by generating an additional model with both NH$_3$-hydrates and a small percentage of crystalline NH$_3$ ice. With the exception of the residual noise spike at 2.207 \um from incomplete solar line cancellation (\S \ref{sec:contam}), this model visually matches the weak absorption bands at 2.2 \um and 2.24 \um more closely. This 'mixed NH$_3$' model has effectively the same $\chi_{\nu}^{2}$ value as the amorphous NH$_3$ model.

\section{Discussion} \label{sec:discussion} 

Our band analysis procedures did not detect any evidence for the presence of crystalline CO$_2$ ice on Miranda's surface, unlike its high abundance on the neighboring moon Ariel. Even in the high S/N spectra from GNIRS and the grand-average spectra from TripleSpec combining multiple nights of data, the three prominent CO$_2$ ice bands between 1.9 and 2.1 \um are not present (Figure \ref{fig:hapke22zoomCO2}). Additionally, we detected essentially no evidence for an absorption feature near 2.13 \um that might have hinted at the presence of CO$_2$ in a molecular mixture. In contrast, a subtle 2.2-\um band is exhibited by many of the Miranda spectra we analyzed.

\subsection{\texorpdfstring{CO$_2$}{CO2} ice}
The prevailing hypothesis for the origin of CO$_2$ ice on the other classical Uranian moons is a radiolytic production mechanism (\S \ref{ssec:CO2bkg}). This hypothesis states that the observed CO$_2$ is produced \textit{in situ} via irradiation of surface carbon compounds and H$_2$O ice, which recombine to form CO$_2$ and other products. Sublimation, sputtering, and other processes cause the CO$_2$ molecules to `hop' in suborbital trajectories across the surface until they either escape or land in a cold trap. Because of the large obliquity of the Uranian system, the parts of the surface that receive the lowest amount of time-averaged sunlight over the Uranian year are at low latitudes. Crystalline CO$_2$ ice is bright and highly reflective, decreasing absorbed solar energy and likely reducing sublimation of CO$_2$ molecules from cold traps once they have been established. Theoretically, the trailing hemispheres would accumulate more CO$_2$ ice because the magnetosphere of Uranus rotates faster than the moons orbit, and might preferentially irradiate their trailing hemispheres and produce CO$_2$ molecules, thereby explaining the observed distribution of CO$_2$ on Ariel, Umbriel, Titania, and Oberon \citep{Grundy2003,Grundy2006,Cart2015,Cart2022}. However, the exact nature of interactions between Uranus' offset and tilted magnetosphere and the surfaces of the moons are still poorly understood \citep[e.g.][]{Kollmann2020}. 

In contrast, this lack of CO$_2$ ice on Miranda, where radiolytic production of CO$_2$ molecules should be possible, could result from its small mass and lower efficiency of retaining volatiles like CO$_2$. Previous works have extensively discussed potential mechanisms for production and destruction of CO$_2$ on the Uranian satellites, such as photolysis or sputtering by charged particles \citep{Grundy2006,Cart2015,Sori2017,Steckloff2022}. Most importantly, a large fraction of the expected thermal velocity distribution of sputtered or sublimated CO$_2$ molecules is greater than the escape velocity of Miranda (193 m s$^{-1}$). \citet{Sori2017} found that approximately half of all CO$_2$ molecules sublimated on Miranda would escape before completing a single suborbital `hop'. \citet{Steckloff2022} found a similar result, adding that the time required for these CO$_2$ molecules to migrate to the antipode of their initial production site on Miranda ($\sim$14 hours) is more than an order of magnitude longer than their expected residence time in the exosphere before escape (approximately one hour). Other processes that could destroy CO$_2$ ice are effectively irrelevant on Miranda, given the rate at which any produced or otherwise mobilized CO$_2$ molecules would be lost. Miranda cannot retain significant CO$_2$ ice deposits with such a large loss fraction. It is therefore not surprising that we do not observe any CO$_2$ ice deposits on Miranda. 

However, given the assumption of ongoing radiolytic production, it is still possible that CO$_2$ is present in Miranda's regolith, perhaps as molecules mixed with H$_2$O ice. This motivated our search for a 2.13-\um absorption band, attributed to the forbidden $2\nu_3$ first overtone of the strong 4.27-\um $\nu_3$ asymmetric stretch band of CO$_2$. This forbidden feature is extremely weak in spectra of pure CO$_2$ ice, but becomes apparent when CO$_2$ is in a molecular mixture with other ices, including H$_2$O and CH$_3$OH \citep{Bernstein2005}. This overtone can vary significantly in its depth and shape in laboratory spectra. In many of our high S/N spectra, visual inspection suggested a weak feature centered at slightly shorter wavelengths ($\sim$2.130 \micron), reminiscent of a weak 2.13-\um band detected on Ariel using SpeX and TripleSpec data \citep{Cart2022}, also centered at $\sim$2.130 \um and spanning 2.123 -- 2.137 \micron. If a molecular mixture including CO$_2$ ice is responsible for this band on Ariel, then it could also be responsible for a similar band on Miranda. However, our band analyses did not find convincing evidence for the 2.13-\um band on Miranda. If CO$_2$ is being produced on Miranda, it is not being trapped in the regolith in sufficient quantities to manifest the 2.13-\um band at a strength detectable in our spectra. 

\subsection{\texorpdfstring{NH$_3$}{NH3}-bearing species\label{ssec:nh3discussion}}
\begin{figure}[ht!]
\centering
\makebox[\textwidth][c]{\includegraphics[width=\textwidth]{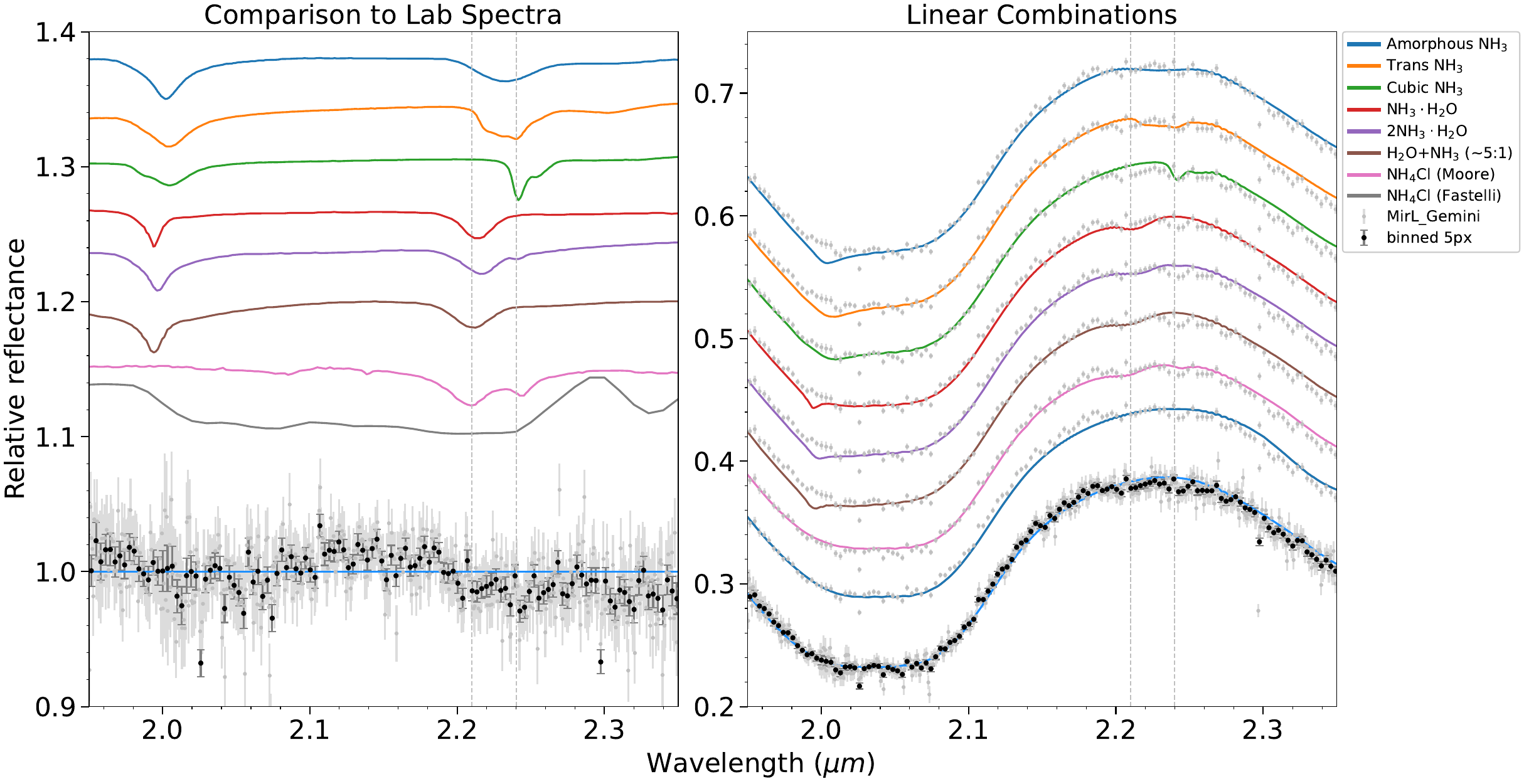}}
\vspace{-15pt}
\caption{\footnotesize (Left panel): The UT201008 GNIRS spectrum of Miranda's leading hemisphere (gray errorbars native resolution, black errorbars binned by 5 pixels), compared to several laboratory-measured absorbance spectra of NH-bearing species (digitized from Figures 6 and 14 of \citet{Moore2007}).  The laboratory spectra are arbitrarily scaled and offset. We also include a reflectance spectrum of NH$_4$Cl at 90 K from \citet{Fastelli2022}. (Right panel): Arbitrarily scaled linear combinations of the lab spectra with our synthetic base model for qualitative comparison of band shapes and locations. Vertical lines are placed at 2.21 and 2.24 \micron.}
\vspace{-5pt}
\label{fig:22umMoorecomp}
\end{figure}

\begin{deluxetable}{cccc}
\tabletypesize{\footnotesize}
\tablecaption{$\chi_{\nu}^{2}$ values of different comparison spectra \label{tab:chi2spectra}}
\tablehead{\colhead{Short name} & \colhead{Figure} & \colhead{Spectrum type} &  \colhead{$\chi_{\nu}^{2}$} }
\startdata
Base model & \ref{fig:22umhapkeices},\ref{fig:hapke22zoomCO2},\ref{fig:22umMoorecomp},\ref{fig:Fastellicomp} & Hapke-Mie & 1.050 \\
10\% illite & \ref{fig:22umhapkeices} & Hapke-Mie & 2.762 \\
5\% kaolinite & \ref{fig:22umhapkeices} & Hapke-Mie & 1.202 \\
2\% propionitrile & \ref{fig:22umhapkeices} & Hapke-Mie & 1.351 \\
2\% ethylamine & \ref{fig:22umhapkeices} & Hapke-Mie & 1.252 \\
2\% methylamine & \ref{fig:22umhapkeices} & Hapke-Mie & 1.236 \\
2\% amorphous NH$_3$ & \ref{fig:22umhapkeices},\ref{fig:hapke22zoomCO2} & Hapke-Mie & 0.938 \\
1\% ethylene & \ref{fig:22umhapkeices} & Hapke-Mie & 1.212 \\
1\% NH$_3$ & \ref{fig:22umhapkeices},\ref{fig:hapke22zoomCO2} & Hapke-Mie & 0.908 \\
1\% NH$_3\cdot$H$_2$O & \ref{fig:22umhapkeices},\ref{fig:hapke22zoomCO2} & Hapke-Mie & 1.051 \\
3\% NH$_3\cdot$H$_2$O & \ref{fig:22umhapkeices} & Hapke-Mie & 1.602 \\
Mixed NH$_3$ & \ref{fig:22umhapkeices},\ref{fig:hapke22zoomCO2} & Hapke-Mie & 0.939 \\\tableline
Amorphous NH$_3$ & \ref{fig:22umMoorecomp} & base + lab absorbance & 0.824 \\
Transition-phase NH$_3$ & \ref{fig:22umMoorecomp} & base + lab absorbance & 0.884 \\
Cubic NH$_3$ & \ref{fig:22umMoorecomp} & base + lab absorbance & 1.065 \\
NH$_3\cdot$H$_2$O & \ref{fig:22umMoorecomp} & base + lab absorbance & 1.014 \\
2NH$_3\cdot$H$_2$O & \ref{fig:22umMoorecomp} & base + lab absorbance & 0.928 \\
NH$_3$+H$_2$O ($\sim$5:1) & \ref{fig:22umMoorecomp} & base + lab absorbance & 0.959 \\
Moore NH$_4$Cl & \ref{fig:22umMoorecomp} & base + lab absorbance & 0.952 \\
Fastelli NH$_4$Cl & \ref{fig:22umMoorecomp},\ref{fig:Fastellicomp} & base + lab reflectance & 73.222 \\
\enddata
\tablecomments{\footnotesize Comparison of $\chi_{\nu}^{2}$ values of our constructed comparison spectra in various figures (measured between the models and unbinned data between 2.15--2.29 \micron). For details on the Hapke-Mie models, see Table \ref{tab:synspectra}. The linear combinations of the Hapke-Mie base model and the laboratory absorbance spectra are arbitrarily scaled and are not quantitatively comparable to the other Hapke-Mie synthetic spectra.}
\end{deluxetable}

In Figure \ref{fig:22umMoorecomp}, we utilized the laboratory absorption spectra of \citet{Moore2007} to investigate the potential contributions of NH$_3$ species. Moore et al. measured the absorption spectra of multiple phases of pure NH$_3$ ice, several NH$_3$-H$_2$O mixtures, two phases of NH$_3$-hydrates, and NH$_4$Cl (from \citet{Moore2003}). We also constructed qualitative linear mixtures using these absorption spectra to qualitatively compare band shapes and locations (right panel of Figure \ref{fig:22umMoorecomp}). These are not physically accurate spectral models, but provide a general idea of how low-level absorption from these features might appear. Quantitative spectral modeling would require the optical constants of these species to be measured, and optical constants have only been published for amorphous NH$_3$ ice, crystalline NH$_3$ ice, and an NH$_3$-hydrate of uncertain stoichiometry (Figures \ref{fig:22umhapkeices},\ref{fig:hapke22zoomCO2}).

Pure NH$_3$ ices of any phase produce bands at longer wavelengths than the broad 2.2-\um band in the Miranda spectrum, but could contribute to shallow absorptions at 2.24 \micron. The NH$_3$-H$_2$O hydrates and NH$_3$-H$_2$O mixture produce a broad band centered between $\sim$2.21 -- 2.22 \micron. Both 2NH$_3\cdot$H$_2$O and the absorbance spectrum of NH$_4$Cl produces a double-band structure with a 2.21-\um and 2.24-\um band. From visual inspection, the NH$_3$-hydrates and NH$_3$-H$_2$O mixtures provide acceptable matches, but NH$_4$Cl (Moore) seems to match the shape and position of the weak 2.2-\um bands in the Miranda spectrum best. When we compare the calculated $\chi_{\nu}^{2}$ values of these linear combinations to the spectrum (Table \ref{tab:chi2spectra}), all but the cubic NH$_3$ mixture have improved $\chi_{\nu}^{2}$ values over the base model ($\chi_{\nu}^{2} = $1.050). As with the synthetic Hapke-Mie spectra, amorphous NH$_3$ ice has the best $\chi_{\nu}^{2}$, but amorphous and transition-phase NH$_3$ ice do not produce band shapes that visually match the Miranda spectrum, and their presence on Miranda is unlikely (\S \ref{sec:modelresults}). The cubic NH$_3$ ice alone produces a visual fit to the 2.24-\um band, but not the band between 2.18 -- 2.23 \micron. Conversely, the 2.18 -- 2.23-\um band is well fit by the NH$_3$-hydrates and NH$_3$-H$_2$O mixtures. The 2NH$_3\cdot$H$_2$O and Moore NH$_4$Cl spectra produce bands at both 2.18 -- 2.23 \um and 2.24 \micron, more consistent with the observed Miranda spectrum. When excluding the amorphous and transition-phase NH$_3$ ices, then 2NH$_3\cdot$H$_2$O, the Moore NH$_4$Cl spectrum, and the NH$_3$-H$_2$O mixture have the best $\chi_{\nu}^{2}$.

However, all of the Moore et al. lab spectra of NH$_3$-bearing species show the 2.0-\um band of NH$_3$ ice, which is not visible in our spectra. This 2.0-\um absorption band is rarely detected on icy bodies. Even in the case of Charon, where the 2.21-\um band is obvious, the presence of a complementary 2.0-\um band is uncertain \citep{Cook2018,Cook2023}. The 2.0-\um band is difficult to detect from ground-based spectra, as it lies in the spectral noise from a deep telluric CO$_2$ band, and even in spectra from New Horizons, the 2.0-\um band is not as strong as would be expected from laboratory spectra \citep{Cook2018,Protopapa2021}. The strong H$_2$O absorption band at 2.0 \um further means that light penetrates much shallower into the regolith ($\sim$ 0.1 mm at 2.0 \micron) than it does at 2.24 \um ($\sim$ 1.6 mm), where H$_2$O only contributes weak absorption and the optical path length is therefore longer \citep{Cart2023}. NH$_3$-bearing species could be present in the subsurface, where they only appear in absorption in regions of the spectrum where the optical path length is long enough to reach them, contributing to the difficulty of detecting the NH$_3$ 2.0-\um absorption feature, or an ammonium salt such as NH$_4$Cl that lacks a 2.0-\um band could be responsible (next section). We also note that these qualitative linear combinations do not incorporate effects like multiple scattering that would make a difference in the strength of the 2.0-\um band. This is demonstrated by the difference in the strength of the 2.0-\um band in the Hapke-Mie synthetic spectra with NH$_3$ ice (Figure \ref{fig:22umhapkeices}), in which the band is much less prominent than in our linear combinations.

The `survivability' of NH$_3$-bearing species is also an open question. As discussed previously, NH$_3$-bearing species exposed at the surface of icy bodies are expected to be dissociated by particle irradiation (such as cosmic rays) on geologically short timescales. Miranda is subject to additional proton and electron bombardment from charged particles trapped in Uranus's magnetosphere, and \citet{Moore2007} estimate timescales for destruction of NH$_3$ ice at Miranda's surface as short as $\sim 10^6$ years. However, dissociated NH$_3$ does not simply disappear. The radiation products can readily recombine into similar forms, such as NH$_4^+$ ions, which can react with other compounds to form NH$_4$-bearing salts.  \citet{Protopapa2021} provides a lengthy review of the mechanisms that could destroy NH$_3$ on Charon. While the radiation environment on Miranda is more hostile to NH$_3$, NH$_4$-bearing species still possess a 2.2-\um band, and NH$_4$-bearing salts may be more resistant to destruction by irradiation.

\subsection{\texorpdfstring{NH$_4$}{NH4}-bearing species and other refractory compounds \label{ssec:nh4discussion}}
\begin{figure}[ht!]
\centering
\makebox[\textwidth][c]{\includegraphics[width=\textwidth]{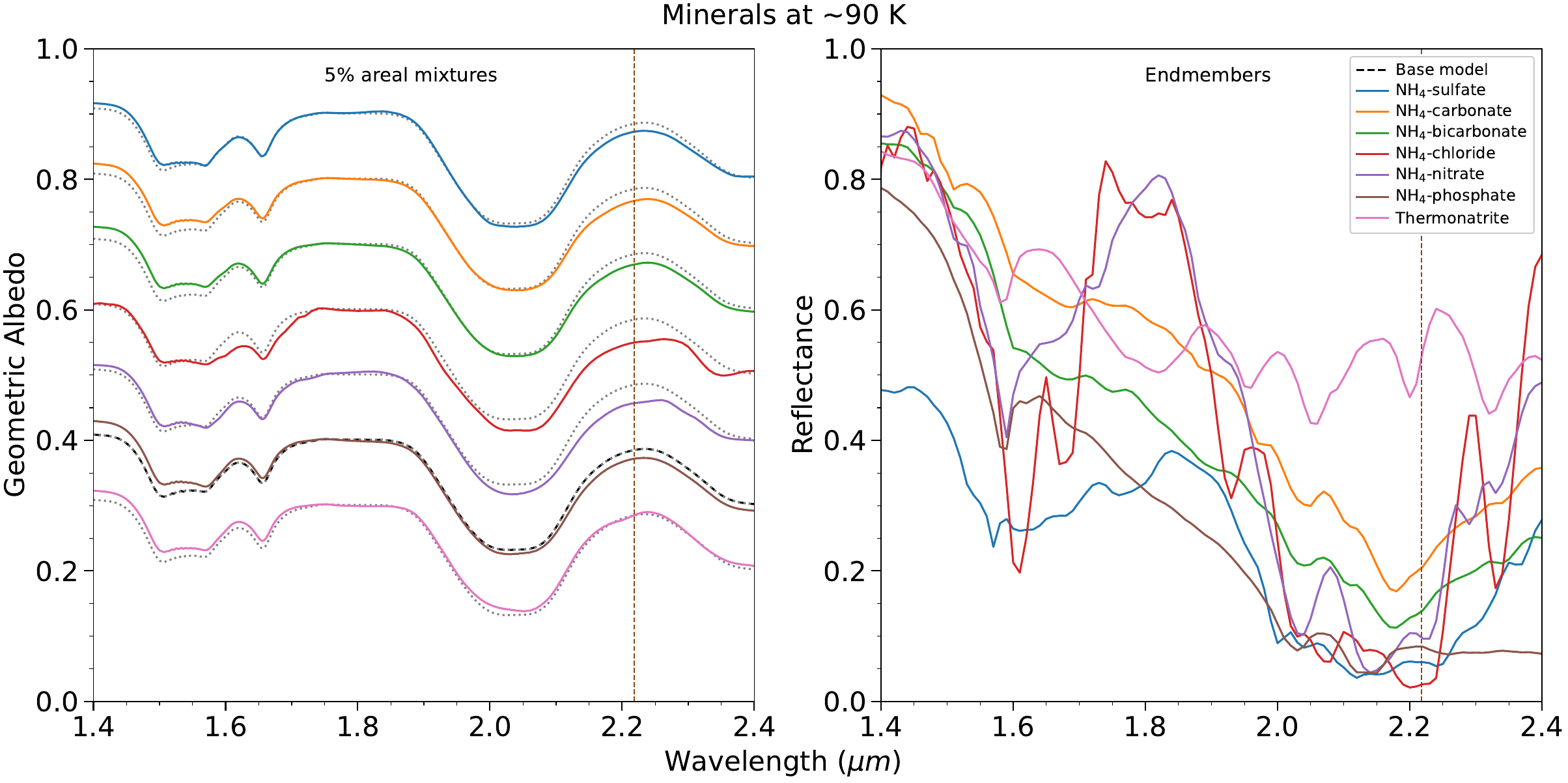}}
\vspace{-15pt}
\caption{\footnotesize (Left panel): Our base model of Miranda's spectrum compared to 5\% areal mixtures of NH$_4$-bearing minerals and thermonatrite. The mixture spectra were normalized to match the base model at 1.754 \micron, then offset in increments of 0.1. The vertical line marks the mean wavelength of our 2.2-\um band measurements at 2.218 \micron. (Right panel): NH$_4$-bearing mineral reflectance spectra measured at 90 K from \citet{Fastelli2022} and thermonatrite measured at 93 K from \citet{DeAngelis2019}.}
\vspace{-5pt}
\label{fig:Fastellicomp}
\end{figure}

Ammonium salts such as NH$_4$Cl have been proposed as a candidate species partially or wholly responsible for the 2.2-\um absorption band on Charon and on other bodies in the Pluto system \citep{Moore2007,Cook2007,Cook2018,Cook2023}. NH$_4$Cl has also been reported in the bright emplaced material in Occator Crater on Ceres \citep{Raponi2019}.
Other NH$_4$-bearing species, such as ammonium carbonate ((NH$_4$)$_2$CO$_3$) or ammonium bicarbonate (NH$_4$HCO$_3$), have also been implicated in the presence of a 2.2-\um absorption \citep{DeSanctis2016,Cart2020Ariel}. The NH$_4^+$ ion is a byproduct of irradiation of NH$_3$-H$_2$O mixtures \citep{Moore2007}, and given other suitable species to react with, it could form NH$_4$-bearing salts. In Figure \ref{fig:Fastellicomp}, we plot reflectance spectra of NH$_4$-bearing species measured at 90 K from \citet{Fastelli2022}, and constructed linear (areal) mixture models of our synthetic base model spectrum, plus a 5\% abundance of various species. The absorption bands of most of these species, with the possible exception of NH$_4$Cl, are at too short wavelengths to account for the 2.2-\um band we see on Miranda, but could contribute to low-level absorption between 2.18 -- 2.25 \micron. As seen in the figure, their broad absorption features tend to simply depress the 2.2-\um continuum of H$_2$O ice relative to the 1.8-\um continuum. Miranda's spectrum exhibits the same characteristic, but the broad nature of these NH$_4$ absorptions observed in reflectance generally limits the ability to identify them, as a depressed 2.2-\um continuum could also result from sub-micron H$_2$O ice grains (\S \ref{ssec:hapkeintro}) or any other compound with a strong blue slope between 1.8 and 2.2 \micron.

However, linear (areal) mixture modeling comes with additional caveats. Linear mixture modeling does not incorporate the nonlinear effects of multiple scattering (important for bright icy surfaces) or other factors in the reflectance of a particular component, such as grain size. These effects are treated in intimate mixture models, but an intimate mixture model requires the optical constants of the material to be measured, and no optical constants for these NH$_4$-bearing species have been published in the scientific literature. Furthermore, direct comparison demonstrates that the reflectance spectrum of a pure compound is distinctly different from the absorption spectrum; compare and contrast the absorbance spectra and reflectance spectra of NH$_4$Cl in Figure \ref{fig:22umMoorecomp}. Although the spectra are arbitrarily scaled, the difference in depths and shapes of the absorption bands are clearly apparent. The Fastelli et al. sample studied in reflectance is much thicker ($\sim$ mm) than the Moore et al. sample studied in absorbance (tens of microns), which makes weaker features easier to detect, but produces saturation and blending of the stronger absorption features. The relative band strengths are different between the Moore and Fastelli NH$_4$Cl spectra: the weak absorptions between 2.0 and 2.2 \um in the Moore spectrum are heavily blended in the Fastelli spectrum, and the stronger absorptions at 2.21 and 2.24-\um are effectively saturated. The Fastelli spectrum also shows a clear band at 2.34-\um that is shallow in the Moore spectrum and absent in the Miranda spectra. Finally, the spectral resolution of the Fastelli spectra are generally inadequate compared to our Miranda spectra and the Moore spectrum.

While our spectra of Miranda were also observed in reflectance, an intimate mixture of small percentages of NH$_4$Cl or another NH$_4$-bearing species in H$_2$O ice could produce a spectral signature more similar to the Moore et al. absorption spectrum. We also note that unlike the rest of the NH$_3$-bearing species in Figure \ref{fig:22umMoorecomp}, NH$_4$Cl lacks a strong 2.0-\um band, consistent with the lack of the 2.0-\um band in our spectra of Miranda.
Confident quantitative estimates are difficult given the weak 2.2-\um absorption, the limited nature of the available laboratory data, and the lack of optical constants for candidate constituents. We suggest that an NH$_4$-bearing species, perhaps NH$_4$Cl, may be partially or wholly responsible for the 2.2-\um band, although NH$_3$-bearing species could contribute to absorption in this range. 

The potential sources (and potential destruction) of NH$_4$Cl on Charon is discussed thoroughly in \citet{Cook2023}. Many of these arguments are also applicable at some level to the Uranian satellites, including Miranda. Chlorine could be delivered to the surface via impacts, or perhaps an endogenic process like cryovolcanism emplaced NaCl-bearing or NH$_4$Cl-bearing brines and/or NH$_3$-rich liquids onto the surface. Some form of this process, likely impact-induced, appears to have occurred at Ceres in Occator Crater \citep{Raponi2019,DeSanctis2020}. Furthermore, carbonaceous material and H$_2$O are readily available at Miranda's surface and particle irradiation is present to dissociate them. In addition to the previously-discussed CO$_2$, if Miranda's surface was supplied with NH$_3$- or NH$_4$-bearing species, NH$_4$-bearing carbonates could be a common byproduct. 

With previous results on Umbriel in mind \citep{Cart2023}, we also generated a synthetic intimate mixture model incorporating the aluminum-bearing phyllosilicates illite and kaolinite (Figure \ref{fig:22umhapkeices}), and an areal mixture model incorporating the hydrated Na-bearing carbonate thermonatrite (Figure \ref{fig:Fastellicomp}). Thermonatrite's 2.2-\um band is centered at shorter wavelengths than our mean band center (2.201 versus 2.218 \micron), but could be contributing to wider low-level absorption between 2.18 -- 2.25 \um (Figure \ref{fig:CO2NH3vstack}). Kaolinite has a blue slope at wavelengths longer than 1.8 \um and a double-band absorption at 2.16 and 2.206 \um that could produce a similar shoulder feature, but as discussed in \S \ref{sec:modelresults}, the only other strong indicator of phyllosilicates in the near-IR ($<$2.5 \micron) is generally the 1.4-\um band, which is difficult to observe from Earth due to telluric absorption.

\section{Conclusions} \label{sec:conclusion}

We conclude that there is no crystalline CO$_2$ ice concentrated in discrete deposits on Miranda's surface, consistent with the explanation that Miranda's weak surface gravity allows CO$_2$ molecules to efficiently escape before being cold trapped. Similarly, we detected no convincing evidence of a 2.13-\um band that could be associated with a molecular mixture of CO$_2$ in H$_2$O or CH$_3$OH ice.

In contrast, we detected a 2.2-\um band at $>2\sigma$ significance in several of our spectra. We do not see evidence for longitudinal trends or hemispherical asymmetries in the distribution of the 2.2-\um band. We compared a high S/N GNIRS spectrum of the leading hemisphere to synthetic spectra and laboratory spectra of several candidate species, including NH-bearing species, nitrogen-bearing organics, and phyllosilicates. We found that the 2.2-\um feature can be best explained by either a combination of NH$_3$-bearing species (e.g. NH$_3$-hydrates and NH$_3$ ice) or by an NH$_4$-bearing salt like NH$_4$Cl, but NH$_4$-carbonates or certain phyllosilicates like kaolinite could contribute to broad and shallow absorption between 2.18 -- 2.25 \micron.

However, the study of NH$_3$-bearing and NH$_4$-bearing species is substantially limited by the available laboratory data. Optical constants of candidate species are few and far between, and reflectance data are often acquired at room temperature and with insufficient spectral resolution to properly capture the blended spectral bands that change position and strength with temperature and composition. Common minerals often show fine structure in absorption bands that are not detected in low resolution reflectance spectra \citep{Clark1990Minerals}. Further laboratory work to determine optical constants of NH$_3$-hydrates, NH$_3$-H$_2$O mixtures, and NH$_4$-bearing salts at high resolution and cryogenic temperatures in the near- to mid-IR should be a high priority for the icy satellites community \citep{Dalton2010}. The results of such laboratory work (spectra and optical constants) must be made publicly downloadable in a data repository for current and future scientists \citep{Roser2021OptConstWhitePaper}.

Future investigations of CO$_2$ ice and NH$_3$-bearing species on Miranda could be carried out with JWST, as the excellent sensitivity of a 6.5-meter IR-optimized space telescope is hard to match even with larger ground-based telescopes. JWST has the ability to investigate the 4.27-\um CO$_2$ ice $\nu_3$ fundamental absorption band, which lies within a wavelength range in which the Earth's atmosphere is completely opaque. This fundamental absorption band is a factor of $\sim$10$^3$ stronger than the 2-\um overtone bands, and the lack of the 2.13-\um band suggests that JWST observations of the 4.27-\um band would be more suitable for further constraining CO$_2$ ice on Miranda and the other Uranian moons. NH$_3$-bearing species would be detectable by JWST with the NH$_3$ $\nu_3$ fundamental band (2.96-\micron) and the hydrated nitrogen (OH-N) features (3.05-\um and 3.1 -- 3.2 \micron) overprinted on the wide 3.0-\um H$_2$O ice band. These bands are far stronger than the weak overtone bands in the near-IR shortward of 2.5 \micron, and different NH-bearing compounds and hydration states are much easier to distinguish from each other at longer wavelengths \citep{Moore2007}. Similarly, NH$_4$-bearing minerals can be detected through features between 3.0 -- 3.5 \um \citep{Bishop2002,Berg2016}.

Spatially resolved spectroscopy of ices on Miranda and the other Uranian satellites would be most effectively carried out by a Uranus orbiter equipped with a NIR imaging spectrometer, analogous to VIMS on Cassini or MISE on the upcoming Europa Clipper mission. The Uranus Orbiter and Probe was the highest priority new Flagship mission recommended by the 2023-2032 Decadal Survey. Such a mission would be able to revolutionize our knowledge of the entire Uranian system, including the icy satellites, and could investigate spatial and spectral variations in surface composition in far more detail than could ever be achieved from Earth.

\begin{acknowledgments}
We wish to extend our gratitude to the observing, engineering, and administrative staff at Apache Point Observatory, Gemini North, and the IRTF, without whom this project would not have been possible, and to the anonymous reviewers whose comments improved this manuscript. 

This work is funded under NASA FINESST grant 80NSSC20K1378, and parts of it were previously funded under the NMSU Astronomy Department's William Webber Voyager Graduate Fellowship. This work is partially based on observations obtained with the ARC 3.5-meter telescope at Apache Point Observatory, which is owned and operated by the Astrophysical Research Consortium. This work also incorporates observations previously obtained at the Infrared Telescope Facility (IRTF), which is operated by the University of Hawaii under contract 80HQTR19D0030 with the National Aeronautics and Space Administration.

Finally, this work is also based in part on observations obtained under program IDs GN-2020B-FT-205 and GN-2021B-FT-210 at the international Gemini Observatory, a program of NSF’s NOIRLab, which is managed by the Association of Universities for Research in Astronomy (AURA) under a cooperative agreement with the National Science Foundation on behalf of the Gemini Observatory partnership: the National Science Foundation (United States), National Research Council (Canada), Agencia Nacional de Investigaci\'{o}n y Desarrollo (Chile), Ministerio de Ciencia, Tecnolog\'{i}a e Innovaci\'{o}n (Argentina), Minist\'{e}rio da Ci\^{e}ncia, Tecnologia, Inova\c{c}\~{o}es e Comunica\c{c}\~{o}es (Brazil), and Korea Astronomy and Space Science Institute (Republic of Korea). The Gemini North and IRTF telescopes are located within the Maunakea Science Reserve and adjacent to the summit of Maunakea. We are grateful for the privilege of observing the Universe from a place that is unique in both its astronomical quality and its cultural significance, and wish to emphasize our respect for the Native Hawaiian community's cultural and historical ties to Maunakea.

This research has made use of the SIMBAD database, operated at CDS, Strasbourg, France.

\end{acknowledgments}

\vspace{5mm}
\facilities{ARC(TripleSpec), Gemini:Gillett(GNIRS), IRTF(SpeX)}

\software{AstroPy \citep{Astropy2013}, NumPy \citep{numpy}, SciPy \citep{scipy}, Matplotlib \citep{Matplotlib}, JPL Horizons Online Ephemeris Service (\url{https://ssd.jpl.nasa.gov/horizons/}), Spextool \citep{Cushing2004,Vacca2003}, SIMBAD \citep{SIMBAD}, SpectRes \citep{Spectres}}

\appendix
\restartappendixnumbering

\section{Spectral contamination \label{sec:contam}}

\begin{figure}[ht!]
\centering
\makebox[\textwidth][c]{\includegraphics[width=\textwidth]{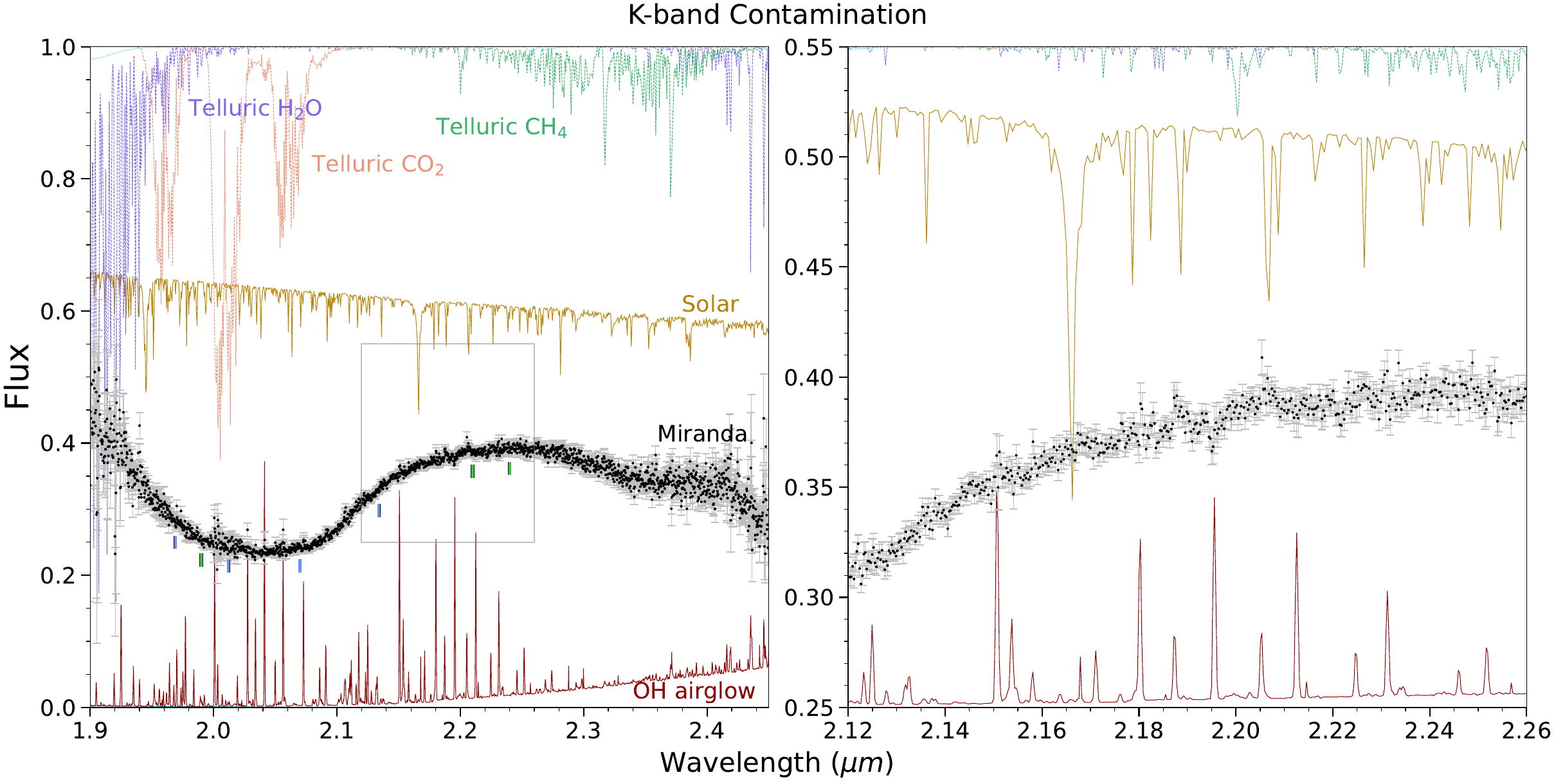}}
\vspace{-15pt}
\caption{\footnotesize (Left panel): Sources of spectral contamination in K-band. The atmospheric transmission curves for telluric species are plotted as the fraction of light transmitted and have correct scaling relative to each other. The solar spectrum has been divided by a 5777 K blackbody spectrum. The plotted Miranda spectrum is a grand average of all TripleSpec spectra. The scaling on the solar, airglow, and Miranda spectrum are arbitrary for visualization purposes. The wavelengths of the absorption features of interest in this work are annotated below the Miranda spectrum: the CO$_2$ ice triplet and the 2.13-\um band are marked in blue, and features associated with NH$_3$-bearing species are marked in green. (Right panel): Inset zoom for the box marked in the left panel. Note the 'spike' residuals in the Miranda spectrum that closely match OH and solar lines.}
\vspace{-5pt}
\label{fig:KbandCont}
\end{figure}

Potential absorption features in the spectrum of Miranda in K-band are also subject to low-level contamination from various sources (Figure \ref{fig:KbandCont}). Absorption from gases in Earth's atmosphere (telluric absorption) is a significant concern for ground-based observations in the near-infrared. Atmospheric H$_2$O vapor is often the most difficult species to correct for, as it has a multitude of strong, narrow absorption lines, and precipitable water vapor along the line of sight can vary locally on short spatial and temporal scales. However, absorption from atmospheric H$_2$O is not a major concern between 2.00 -- 2.40 \micron, which encompasses almost all of the absorption features of interest in this work.

Telluric CO$_2$ has three strong overtone bands which are the major source of atmospheric absorption between 1.95 -- 2.10 \micron. Unlike H$_2$O absorption, which varies strongly due to local weather conditions, CO$_2$ is well-mixed in the Earth's atmosphere and variations in absorption strength are largely based on optical path length through the atmosphere (airmass). However, fine structure in the deepest parts of the CO$_2$ overtone bands are partially resolved in our highest-resolution data (R$\sim$3500), and telluric correction procedures often leave high-frequency residual noise in these regions. This is visible in many of the Miranda spectra as an increased scatter in the data points between 2.00 -- 2.02 \micron, and to a lesser extent between 1.95 -- 1.97 and 2.05 -- 2.07 \micron. These wavelengths ranges overlap the narrow CO$_2$ ice bands, but this should not prevent the detection of CO$_2$ ice given the high S/N of this dataset.

The 2.2-\um region experiences weak absorption from narrow atmospheric CH$_4$ bands, with a somewhat stronger feature at 2.200-\micron. However, like CO$_2$, telluric CH$_4$ is generally well-mixed in the Earth's atmosphere. We find that the usual telluric standard procedures are effective at correcting absorption features from atmospheric CH$_4$, as evidenced by the lack of residuals from the other CH$_4$ absorption lines in K-band, so we do not expect Miranda's 2.2-\um feature to be contaminated to a substantial degree by atmospheric CH$_4$. Atmospheric CO$_2$ and H$_2$O do not have significant absorption features between 2.10 -- 2.35 \micron.

OH airglow emission lines are generally well-corrected in our spectra, but stacking many spectra can leave small positive or negative residual features and increased error estimates on certain data points, such as at 2.041, 2.150, and 2.195 \micron. The upwards slope of the continuum in the plotted OH spectrum at longer wavelengths ($>$2.1 \micron) is blackbody radiation from the telescope and atmosphere. This slope is subtracted from the science data during data reduction procedures and does not factor into the final spectra.

We also note that there are mismatches in the strength of stellar absorption features between the G-type telluric standard stars and the reflected solar spectrum. When dividing the observed spectra of the object by the standard star, these mismatches appear as narrow residual 'spikes' or 'dips' in the final spectrum. This includes the inconvenient presence of a residual spike at 2.207 \micron, which appears to be due to a mismatch in the strength of a \ion{Si}{1} metal line (see Figure \ref{fig:hapke22zoomCO2}). This identification is supported by other small, but noticeable positive residuals from other \ion{Si}{1} lines in K-band at 2.093, 2.135, and 2.188 \um \citep{Mohler1953}. Larger mismatches are also sometimes apparent in the strength of the \ion{H}{1} Brackett $\gamma$ line at 2.166 \micron, and any apparent absorption features in the range 2.155 -- 2.175 \um should be treated with suspicion. In general, any narrow residual features in the high-resolution Miranda spectra appear to be either due to OH airglow lines or from mismatches in solar/stellar spectra. 

Finally, we note that scattered light from the disk of Uranus is a major source of contamination in spectra of Miranda at wavelengths shorter than $\sim$1.7 \micron. However, Uranus is very faint in K-band due to absorption by CH$_4$ and the H$_2$ pressure-induced dipole \citep{FinkLarson1979,Baines1998}, which includes the wavelength ranges and absorption features of interest in this work. The contamination of the Miranda spectra by the spectrum of Uranus is therefore negligible for the purposes of this study.

\newpage
\bibliography{MirBib}{}
\bibliographystyle{aasjournal}

\end{document}